\documentclass[numbers]{sigplanconf}

\usepackage{amsmath} 
\usepackage{amssymb} 
\usepackage{amsthm} 
\usepackage{mathtools} 
\usepackage{stmaryrd} 
\usepackage{array}    
\usepackage{comment} 
\usepackage{enumitem}       
\usepackage[
  colorlinks=true,
  allcolors=black,
  breaklinks,
  draft=false]{hyperref} 
\usepackage{balance} 

\usepackage{tikz}
\usepackage{pgfplots}
\usetikzlibrary{matrix,arrows}
\newenvironment{customlegend}[1][]{%
    \begingroup
    \csname pgfplots@init@cleared@structures\endcsname
    \pgfplotsset{#1}%
}{%
    \csname pgfplots@createlegend\endcsname
    \endgroup
}%
\def\addlegendimage{\csname pgfplots@addlegendimage\endcsname}

\usepackage{mathpartir} 
\usepackage{galois}     

\newtheorem{lemma}{Lemma}
\newtheorem{theorem}{Theorem}
\usepackage{tikz}
\usepackage{pgfplots}

\newcommand{\shortnearrow}{\mathrel{\rotatebox[origin=c]{33}{$\rightarrow$}}}
\newcommand{\shortsearrow}{\mathrel{\rotatebox[origin=c]{-33}{$\rightarrow$}}}
\newcommand{\trueaf}{\hspace{-.7em}\textsuperscript{\texttt{AGDA}\checkmark}}

\newif\ifsecbookends
\newif\ifsubbookends
\newif\ifparbookends
\newif\ifdetails

\secbookendstrue
\subbookendstrue
\parbookendstrue
\detailstrue

\begin{document}


\conferenceinfo{ICFP'16}{September 18--24, 2016, Nara, Japan}
\copyrightyear{2016}
\copyrightdata{978-1-4503-4219-3/16/09}
\copyrightdoi{2951913.2951934}
\proceedings{Proceedings of the ACM SIGPLAN International
    Conference on Functional Programming}



\title{Constructive Galois Connections} 
\subtitle{Taming the Galois Connection Framework for Mechanized Metatheory}

\authorinfo{David Darais}
           {University of Maryland, USA}
           {darais@cs.umd.edu}

\authorinfo{David Van Horn}
           {University of Maryland, USA}
           {dvanhorn@cs.umd.edu}

\maketitle


\begin{abstract} 

Galois connections are a foundational tool for structuring abstraction in
semantics and their use lies at the heart of the theory of abstract
interpretation. Yet, mechanization of Galois connections remains limited to
restricted modes of use, preventing their general application
in mechanized metatheory and certified programming.

This paper presents \emph{constructive Galois connections}, a variant of
Galois connections that is effective both on paper and in proof assistants;
is complete with respect to a large subset of classical Galois connections;
and enables more general reasoning principles, including the
``calculational'' style advocated by Cousot.

To design constructive Galois connection we identify a restricted
mode of use of classical ones which is both general and amenable to
mechanization in dependently-typed functional programming
languages. Crucial to our metatheory is the addition of
monadic structure to Galois connections to control a ``specification
effect''.  Effectful calculations may reason classically, while pure
calculations have extractable computational content. Explicitly
moving between the worlds of specification and implementation is
enabled by our metatheory.

To validate our approach, we provide two case studies in mechanizing existing
proofs from the literature: one uses calculational abstract interpretation to
design a static analyzer, the other forms a semantic basis for gradual
typing.  Both mechanized proofs closely follow their original
paper-and-pencil counterparts, employ reasoning principles not captured by
previous mechanization approaches, support the extraction of verified
algorithms, and are novel.

\end{abstract} 


\section{Introduction} \label{s:introduction} 

Abstract interpretation is a general theory of sound approximation
widely applied in programming language semantics, formal verification,
and static analysis~\cite{dvanhorn:Cousot1976Static,
  dvanhorn:Cousot:1977:AI, dvanhorn:Cousot1979Systematic,
  dvanhorn:Cousot1992Inductive, dvanhorn:Cousot2014Galois}.  In
abstract interpretation, properties of programs are related between a
pair of partially ordered sets: a concrete domain,
$\langle\mathcal{C},\sqsubseteq\rangle$, and an abstract domain,
$\langle\mathcal{A},\preceq\rangle$.  When concrete properties have a
$\preceq$-most precise abstraction, the correspondence is a
\emph{Galois connection}, formed by a pair of mappings between the
domains known as \emph{abstraction} $\alpha \in \mathcal{C} \mapsto
\mathcal{A}$ and \emph{concretization} $\gamma \in \mathcal{A} \mapsto
\mathcal{C}$ such that $
c \sqsubseteq \gamma(a) \iff \alpha(c) \preceq a$.

Since its introduction by Cousot and Cousot in the late 1970s, this theory has
formed the basis of many static analyzers, type systems, model-checkers,
obfuscators, program transformations, and many more applications~\cite{local:cousot-bib}.

Given the remarkable set of intellectual tools contributed by this theory, an
obvious desire is to incorporate its use into proof assistants to mechanically
verify proofs by abstract interpretation. When embedded in a proof assistant,
verified algorithms such as static analyzers can then be extracted from these
proofs.

Monniaux first achieved the goal of mechanization for the theory of
abstract interpretation with Galois connections in
Coq~\cite{local:monniaux:1998:ms-thesis}.  However, he notes that the
abstraction side ($\alpha$) of Galois connections poses a serious
problem since it requires the admission of non-constructive axioms.
Use of these axioms prevents the extraction of certified programs.  So
while Monniaux was able to mechanically verify proofs by abstract
interpretation in its full generality, certified artifacts could not
generally be extracted.

Pichardie subsequently tackled the extraction problem by mechanizing a
restricted formulation of abstract interpretation that relied only on
the concretization ($\gamma$) side of Galois
connections~\cite{local:pichardie:2005:phd-thesis}.  Doing so avoids
the use of axioms and enables extraction of certified artifacts.
This proof technique is effective and has been used to
construct several certified static
analyzers~\cite{local:pichardie:2005:phd-thesis,
  dvanhorn:Cachera2010Certified, dvanhorn:Blazy2013Formal,
  dvanhorn:Barthe2007Certified}, most notably the Verasco static
analyzer, part of the CompCert C
compiler~\cite{dvanhorn:Jourdan2015FormallyVerified, dvanhorn:Leroy2009Formal}.
Unfortunately, this approach sacrifices the full generality of the theory.
While in principle the technique could achieve mechanization of
existing soundness \emph{theorems}, it cannot do so faithful to
existing \emph{proofs}.  In particular, Pichardie
writes~\cite[p.~55]{local:pichardie:2005:phd-thesis}:\footnote{
  Translated from French by the present authors.}
\begin{quotation}
\noindent
The framework we have retained nevertheless loses an important
property of the standard framework: being able to derive a correct
approximation $f^{\sharp }$ from the specification $\alpha  \circ f \circ \gamma $. Several
examples of such derivations are given by
Cousot~\cite{davdar:cousot:1999:calculational}. It seems interesting to find a
framework for this kind of symbolic manipulation, while remaining
easily formalizable in Coq.
\end{quotation}
This important property is the so-called ``calculational'' style,
whereby an abstract interpreter ($f^{\sharp }$) is derived in a
correct-by-construction manner from a concrete interpreter ($f$)
composed with abstraction and concretization ($\alpha  \circ f \circ \gamma $).
This style of abstract interpretation is detailed in Cousot's
monograph~\cite{davdar:cousot:1999:calculational}, which
concludes:
\begin{quotation}
\noindent
The emphasis in these notes has been on the correctness of the design
by calculus. The mechanized verification of this formal development
using a proof assistant can be foreseen with automatic extraction of a
correct program from its correctness proof.
\end{quotation}
In the subsequent 17 years, this vision has remained unrealized, and clearly
the paramount technical challenge in achieving it is obtaining both
\emph{generality} and \emph{constructivity} in a single framework.

This paper contributes \emph{constructive Galois connections}, a framework for
mechanized abstract interpretation with Galois connections that achieves both
generality and constructivity, thereby enabling calculational style proofs
which make use of both abstraction ($\alpha$) and concretization ($\gamma$),
while also maintaining the ability to extract certified static analyzers.

We develop constructive Galois connections from the insight that many classical
Galois connections used in practice are of a particular restricted form, which
is reminiscent of a direct-style verification. Constructive Galois connections
are the general abstraction theory for this setting and can be mechanized
effectively.

We observe that constructive Galois connections contain monadic structure which
isolates classical specifications from constructive algorithms. Within the
effectful fragment, all of classical Galois connection reasoning can be
employed, while within the pure fragment, functions must carry computational
content. Remarkably, calculations can move between these modalities and
verified programs may be extracted from the end result of calculation.

To support the utility of our theory we build a library for
constructive Galois connections in Agda~\cite{local:norell:thesis} and
mechanize two existing abstract interpretation proofs from the
literature.  The first is drawn from Cousot's
monograph~\cite{davdar:cousot:1999:calculational}, which derives a
correct-by-construction analyzer from a specification induced by a
concrete interpreter and Galois connection.
The second is drawn from Garcia, Clark and Tanter's ``Abstracting
Gradual Typing''~\cite{local:garcia:2016:agt}, which uses abstract
interpretation to derive static and dynamic semantics for gradually
typed languages from traditional static types.
Both proofs use the ``important property of the
standard framework'' identified by Pichardie, which is not handled by prior
mechanization approaches.
The mechanized proofs closely follow the original pencil-and-paper
proofs, which use both abstraction and concretization, while still
enabling the extraction of certified algorithms.  Neither of these
papers have been previously mechanized.  Moreover, we know of no
existing mechanized proof involving calculational abstract
interpretation.

Finally, we develop the metatheory of constructive Galois connections, prove
them sound, and make precise their relationship to classical Galois
connections. The metatheory is itself mechanized; claims are marked with
``\texttt{AGDA}\checkmark'' whenever they are proved in Agda. (All claims are marked.)

\paragraph{Contributions} 
This paper contributes the following:
\begin{itemize}
\item a foundational theory of constructive Galois connections which is both
  general and amenable to mechanization using a dependently typed functional
  programming language;
\item a proof library and two case studies from the literature for mechanized
  abstract interpretation; and
\item the first mechanization of calculational abstract interpretation.
\end{itemize}

The remainder of the paper is organized as follows. First we give a tutorial on
verifying a simple analyzer from two different perspectives: direct verification
(\S\ref{s:the-direct-approach}) and abstract interpretation with Galois
connections (\S\ref{s:classical-ai}), highlighting mechanization issues along
the way. We then present constructive Galois connections as a marriage of the
two approaches (\S\ref{s:constructive-gcs}). We provide two case studies: the
mechanization of an abstract interpreter from Cousot's calculational monograph
(\S\ref{s:case-cdgai}), and the mechanization of Garcia, Clark and Tanter's
work on gradual typing \emph{via} abstract interpretation (\S\ref{s:case-agt}).
Finally, we formalize the metatheory of constructive Galois connections
(\S\ref{s:metatheory}), relate our work to the literature (\S\ref{s:related}),
and conclude (\S\ref{s:conclusions}).


\section{Verifying a Simple Static Analyzer} \label{s:tutorial} 

In this section we contrast two perspectives on verifying a static analyzer:
using a direct approach, and using the theory of abstract interpretation with
Galois connections. The direct approach is simple but lacks the benefits of a
general abstraction framework. Abstract interpretation provides these benefits,
but at the cost of added complexity and resistance to mechanized verification.
In Section~\ref{s:constructive-gcs} we present an alternative perspective: abstract
interpretation with \emph{constructive} Galois connections---the topic of this
paper. Constructive Galois connections marry the worlds presented in this
section, providing the simplicity of direct verification, the benefits of a
general abstraction framework, and support for mechanized verification.

To demonstrate both verification perspectives we design a parity analyzer in
each style. For example, a parity analysis discovers that $2$ has parity
$\texttt{\textsc{e}}\texttt{\textsc{v}}\texttt{\textsc{e}}\texttt{\textsc{n}}$, $succ(1)$ has parity $\texttt{\textsc{e}}\texttt{\textsc{v}}\texttt{\textsc{e}}\texttt{\textsc{n}}$, and $n+n$ has parity $\texttt{\textsc{e}}\texttt{\textsc{v}}\texttt{\textsc{e}}\texttt{\textsc{n}}$ if $n$ has
parity $\texttt{\textsc{o}}\texttt{\textsc{d}}\texttt{\textsc{d}}$. Rather than sketch the high-level details of a complete static
analyzer, we instead zoom into the low-level details of a tiny fragment:
analyzing the successor arithmetic operation $succ(n)$. At this level of detail
the differences, advantages and disadvantages of each approach become apparent.

\subsection{The Direct Approach} \label{s:the-direct-approach} 

Using the direct approach to verification one designs the analyzer, defines
what it means for the analyzer to be sound, and then completes a proof of
soundness. Each step is done from scratch, and in the simplest way possible. 

This approach should be familiar to most readers, and exemplifies how most
researchers approach formalizing soundness for static analyzers: first posit
the analyzer and soundness framework, then attempt the proof of soundness. One
limitation of this approach is that the setup---which gives lots of room for
error---isn't known to be correct until after completing the final proof.
However, a benefit of this approach is it can easily be mechanized.

\paragraph{Analyzing Successor}

A parity analysis answers questions like: ``what is the parity of $succ(n)$,
given that $n$ is even?'' To answer these questions, imagine replacing $n$ with
the symbol $\texttt{\textsc{e}}\texttt{\textsc{v}}\texttt{\textsc{e}}\texttt{\textsc{n}}$, a stand-in for an arbitrary even number. This hypothetical
expression $succ(\texttt{\textsc{e}}\texttt{\textsc{v}}\texttt{\textsc{e}}\texttt{\textsc{n}})$ is interpreted by defining a successor function over
parities, rather than numbers, which we call $succ^{\sharp }$. This successor operation
on parities is designed such that if $p$ is the parity for $n$, $succ^{\sharp }(p)$ will
be the parity of $succ(n)$:
\begin{align*}& \mathbb{P} \coloneqq  \{ \texttt{\textsc{e}}\texttt{\textsc{v}}\texttt{\textsc{e}}\texttt{\textsc{n}},\texttt{\textsc{o}}\texttt{\textsc{d}}\texttt{\textsc{d}}\}  && succ^{\sharp }(\texttt{\textsc{e}}\texttt{\textsc{v}}\texttt{\textsc{e}}\texttt{\textsc{n}}) \coloneqq  \texttt{\textsc{o}}\texttt{\textsc{d}}\texttt{\textsc{d}} 
\\& succ^{\sharp } : \mathbb{P} \rightarrow  \mathbb{P}  && succ^{\sharp }(\texttt{\textsc{o}}\texttt{\textsc{d}}\texttt{\textsc{d}}) \coloneqq  \texttt{\textsc{e}}\texttt{\textsc{v}}\texttt{\textsc{e}}\texttt{\textsc{n}}
\end{align*}

\paragraph{Soundness}

The soundness of $succ^{\sharp }$ is defined using an interpretation for parities, which
we notate $\llbracket p\rrbracket $:
\begin{align*}& \llbracket \textunderscore\rrbracket  : \mathbb{P} \rightarrow  \wp (\mathbb{N}) && \begin{aligned} & \llbracket \texttt{\textsc{e}}\texttt{\textsc{v}}\texttt{\textsc{e}}\texttt{\textsc{n}}\rrbracket  \coloneqq  \{  n \mathrel{|} even(n) \} 
                                \\ & \llbracket \texttt{\textsc{o}}\texttt{\textsc{d}}\texttt{\textsc{d}}\rrbracket  \coloneqq  \{  n \mathrel{|} odd(n) \} 
                   \end{aligned}
\end{align*}
Given this interpretation, a parity $p$ is a valid analysis result for a number
$n$ if the interpretation for $p$ contains $n$, that is $n \in  \llbracket p\rrbracket $. The analyzer
$succ^{\sharp }(p)$ is then sound if, when $p$ is a valid analysis result for some
number $n$, $succ^{\sharp }(p)$ is a valid analysis result for $succ(n)$:
\begin{align*}& n \in  \llbracket p\rrbracket  \implies   succ(n) \in  \llbracket succ^{\sharp }(p)\rrbracket  \tag{DA-Snd} \label{e:DA-Snd} \end{align*}
The proof is by case analysis on $\llbracket p\rrbracket $; we show the case $p=\texttt{\textsc{e}}\texttt{\textsc{v}}\texttt{\textsc{e}}\texttt{\textsc{n}}$:
\begin{align*}& n \in  \llbracket \texttt{\textsc{e}}\texttt{\textsc{v}}\texttt{\textsc{e}}\texttt{\textsc{n}}\rrbracket 
\\& \Leftrightarrow  even(n)                 && \lbag \text{ defn. of $\llbracket \textunderscore\rrbracket $ }\rbag 
\\& \Leftrightarrow  odd(succ(n))            && \lbag \text{ defn. of $even/odd$ }\rbag 
\\& \Leftrightarrow  succ(n) \in  \llbracket \texttt{\textsc{o}}\texttt{\textsc{d}}\texttt{\textsc{d}}\rrbracket          && \lbag \text{ defn. of $\llbracket \textunderscore\rrbracket $ }\rbag 
\\& \Leftrightarrow  succ(n) \in  \llbracket succ^{\sharp }(\texttt{\textsc{e}}\texttt{\textsc{v}}\texttt{\textsc{e}}\texttt{\textsc{n}})\rrbracket  && \lbag \text{ defn. of $succ^{\sharp }$ }\rbag 
\end{align*}

\paragraph{An Even Simpler Setup}

There is another way to define and prove soundness: use a function which
computes the parity of a number in the definition of soundness. This approach
is even simpler, and will help foreshadow the constructive Galois connection
setup.
\begin{align*}& parity : \mathbb{N} \rightarrow  \mathbb{P} && \begin{aligned} & parity(0) \coloneqq  \texttt{\textsc{e}}\texttt{\textsc{v}}\texttt{\textsc{e}}\texttt{\textsc{n}}
                                \\ & parity(succ(n)) \coloneqq  flip(parity(n))
                   \end{aligned}
\end{align*}
where $flip(\texttt{\textsc{e}}\texttt{\textsc{v}}\texttt{\textsc{e}}\texttt{\textsc{n}}) \coloneqq  \texttt{\textsc{o}}\texttt{\textsc{d}}\texttt{\textsc{d}}$ and $flip(\texttt{\textsc{o}}\texttt{\textsc{d}}\texttt{\textsc{d}}) \coloneqq  \texttt{\textsc{e}}\texttt{\textsc{v}}\texttt{\textsc{e}}\texttt{\textsc{n}}$. This gives an alternative and
equivalent way to relate a number and a parity, due to the following
correspondence:
\begin{align*}& n \in  \llbracket p\rrbracket  \iff   parity(n) = p \tag{DA-Corr} \label{e:DA-Corr} \end{align*}
The soundness of the analyzer is then restated:
\begin{align*}& parity(n) = p \implies   parity(succ(n)) = succ^{\sharp }(p) \end{align*}
or by substituting $parity(n) = p$:
\begin{align*}& parity(succ(n)) = succ^{\sharp }(parity(n)) \tag{DA-Snd*} \label{e:DA-SndS} \end{align*}
Both this statement for soundness and its proof are simpler than before. The
proof follows directly from the definition of $parity$ and the fact that
$succ^{\sharp }$ is identical to $flip$.

\paragraph{The Main Idea} Correspondences like \eqref{e:DA-Corr}---between an
interpretation for analysis results ($\llbracket p\rrbracket $) and a function which computes
analysis results ($parity(n)$)---are central to the constructive Galois
Connection framework we will describe in Section~\ref{s:constructive-gcs}. Using
correspondences like these, we build a general theory of abstraction that
recovers this direct approach to verification, mirrors all of the benefits of
abstract interpretation with classical Galois connections, supports mechanized
verification, and in some cases simplifies the proof effort. We also observe
that many classical Galois connections used in practice can be ported to this
simpler setting.

\paragraph{Mechanized Verification}

This direct approach to verification is amenable to mechanization using proof
assistants like Coq and Agda. These tools are founded on constructive logic in
part to support verified program extraction. In constructive logic, functions
$f : A \rightarrow  B$ are computable and often defined inductively to ensure they can be
extracted and executed as programs. Analogously, propositions $P : \wp (A)$ are
encoded constructively as undecidable predicates $P : A \rightarrow  prop$ where $x \in  P \Leftrightarrow 
P(x)$.

To mechanize the verification of $succ^{\sharp }$ we first translate its definition to a
constructive setting unmodified. Next we translate $\llbracket p\rrbracket $ to a relation
$I(p,n)$ defined inductively on $n$:
\begin{mathpar} \inferrule
   {\ }
   {I(\texttt{\textsc{e}}\texttt{\textsc{v}}\texttt{\textsc{e}}\texttt{\textsc{n}},0)} 

 \inferrule
   {I(p,n)}
   {I(flip(p),succ(n))} 
\end{mathpar}
The mechanized proof of \eqref{e:DA-Snd} using $I$ is analogous to the one we
sketched, and the mechanized proof of \eqref{e:DA-SndS} follows directly by
computation. The proof term for \eqref{e:DA-SndS} in both Coq and Agda is simply
\texttt{refl}, the reflexivity judgment for syntactic equality modulo
computation in constructive logic.

\paragraph{Wrapping Up}

The two different approaches to verification we present are distinguished by
which parts of the design are postulated, and which parts are derived. Using
the direct approach, the analysis $succ^{\sharp }$, the interpretation for parities
$\llbracket p\rrbracket $ and the definition of soundness are all postulated up-front. When the
soundness setup is correct but the analyzer is wrong, the proof at the end will
not go through and the analyzer must be redesigned. Even worse, when the
soundness setup and the analyzer are both wrong, the proof might actually
succeed, giving a false assurance in the soundness of the analyzer. However,
the direct approach is attractive because it is simple and supports mechanized
verification.


\subsection{Classical Abstract Interpretation} \label{s:classical-ai} 

To verify an analyzer using abstract interpretation with Galois connections,
one first designs \emph{abstraction} and \emph{concretization} mappings between
sets $\mathbb{N}$ and $\mathbb{P}$. These mappings are used to synthesize an optimal
specification for $succ^{\sharp }$. One then proves that a postulated $succ^{\sharp }$ meets this
synthesized specification, or alternatively derives the definition of $succ^{\sharp }$
directly from the optimal specification. 

In contrast to the direct approach, rather than design the definition of
\emph{soundness}, one instead designs the definition of \emph{abstraction}
within a structured framework. Soundness is not designed, it is derived from
the definition of abstraction. Finally, there is added boilerplate in the
abstract interpretation approach, which requires lifting definitions and proofs
to powersets $\wp (\mathbb{N})$ and $\wp (\mathbb{P})$.

\paragraph{Abstracting Sets}

Powersets are introduced in abstraction and concretization functions to support
relational mappings, like mapping the symbol $\texttt{\textsc{e}}\texttt{\textsc{v}}\texttt{\textsc{e}}\texttt{\textsc{n}}$ to the set of all even
numbers. The mappings are therefore between \emph{powersets} $\wp (\mathbb{N})$ and $\wp (\mathbb{P})$.
The abstraction and concretization mappings must also satisfy correctness
criteria, detailed below, at which point they are called a \emph{Galois
connection}.

The abstraction mapping from $\wp (\mathbb{N})$ to $\wp (\mathbb{P})$ is notated $\alpha $, and is defined as
the pointwise lifting of $parity(n)$:
\begin{align*}& \alpha  : \wp (\mathbb{N})\! \rightarrow \! \wp (\mathbb{P}) && \alpha (N) \coloneqq  \{  parity(n) \mathrel{|} n \in  N \} 
\end{align*}
The concretization mapping from $\wp (\mathbb{P})$ to $\wp (\mathbb{N})$ is notated $\gamma $, and is defined
as the flattened pointwise lifting of $\llbracket p\rrbracket $:
\begin{align*}& \gamma  : \wp (\mathbb{P})\! \rightarrow \! \wp (\mathbb{N}) && \gamma (P) \coloneqq  \{  n \mathrel{|} p \in  P \wedge  n \in  \llbracket p\rrbracket  \} 
\end{align*}
The correctness criteria for $\alpha $ and $\gamma $ is the correspondence:
\begin{align*}& N \subseteq  \gamma (P) \iff   \alpha (N) \subseteq  P \tag{GC-Corr} \label{e:GC-Corr} \end{align*}
The correspondence means that, to relate elements of different sets---in this
case $\wp (\mathbb{N})$ and $\wp (\mathbb{P})$---it is equivalent to relate them through either $\alpha $ or
$\gamma $. Mappings like $\alpha $ and $\gamma $ which share this correspondence are called
Galois connections.

An equivalent correspondence to \eqref{e:GC-Corr} is two laws relating compositions
of $\alpha $ and $\gamma $, called \emph{expansive} and \emph{reductive}:
\begin{align*}& N \subseteq  \gamma (\alpha (N))  \tag{GC-Exp} \label{e:GC-Exp}
\\& \alpha (\gamma (P)) \subseteq  P  \tag{GC-Red} \label{e:GC-Red}
\end{align*}
Property \eqref{e:GC-Red} ensures $\alpha $ is the best abstraction possible w.r.t. $\gamma $.
For example, a hypothetical definition $\alpha (N) \coloneqq  \{ \texttt{\textsc{e}}\texttt{\textsc{v}}\texttt{\textsc{e}}\texttt{\textsc{n}},\texttt{\textsc{o}}\texttt{\textsc{d}}\texttt{\textsc{d}}\} $ is expansive but not
reductive because $\alpha (\gamma (\{ \texttt{\textsc{e}}\texttt{\textsc{v}}\texttt{\textsc{e}}\texttt{\textsc{n}}\} )) \not\subseteq  \{ \texttt{\textsc{e}}\texttt{\textsc{v}}\texttt{\textsc{e}}\texttt{\textsc{n}}\} $.

In general, Galois connections are defined for arbitrary posets $\langle A,\sqsubseteq ^A\rangle $ and
$\langle B,\sqsubseteq ^B\rangle $. The correspondence \eqref{e:GC-Corr} and its expansive/reductive variants
are generalized in this setting to use partial orders $\sqsubseteq ^A$ and $\sqsubseteq ^B$ instead of
subset ordering. We are also omitting monotonicity requirements for $\alpha $ and $\gamma $
in our presentation (although \eqref{e:GC-Corr} implies monotonicity).

\paragraph{Powerset Lifting}

The original functions $succ$ and $succ^{\sharp }$ cannot be related through $\alpha $ and $\gamma $
because they are not functions between powersets. To remedy this they are
lifted pointwise:
\begin{align*}& \mathord{\uparrow }succ : \wp (\mathbb{N}) \rightarrow  \wp (\mathbb{N})  && \mathord{\uparrow }succ(N) \coloneqq  \{ succ(n) \mathrel{|} n \in  N\} 
\\& \mathord{\uparrow }succ^{\sharp } : \wp (\mathbb{P}) \rightarrow  \wp (\mathbb{P}) && \mathord{\uparrow }succ^{\sharp }(P) \coloneqq  \{ succ^{\sharp }(p) \mathrel{|} p \in  P\} 
\end{align*}
These lifted operations are called the \emph{concrete interpreter} and
\emph{abstract interpreter}, because the former operates over the
\emph{concrete domain} $\wp (\mathbb{Z})$ and the latter over the \emph{abstract domain}
$\wp (\mathbb{P})$. In the framework of abstract interpretation, static analyzers are
just abstract interpreters. Lifting to powersets is necessary to use the
abstract interpretation framework, and has the negative effect of adding
boilerplate to definitions and proofs of soundness.

\paragraph{Soundness}

The definition of soundness for $succ^{\sharp }$ is synthesized by relating $\mathord{\uparrow }succ^{\sharp }$ to
$\mathord{\uparrow }succ$ composed with $\alpha $ and $\gamma $: 
\begin{align*}& \alpha (\mathord{\uparrow }succ(\gamma (P))) \subseteq  \mathord{\uparrow }succ^{\sharp }(P) \tag{GC-Snd} \label{e:GC-Snd} \end{align*}
The left-hand side of the ordering is an optimal specification for any
abstraction of $\mathord{\uparrow }succ$ (a consequence of \eqref{e:GC-Corr}), and the subset ordering
says $\mathord{\uparrow }succ^{\sharp }$ is an over-approximation of this optimal specification. The
reason to over-approximate is because the specification is a mathematical
description, and the abstract interpreter is usually an algorithm, and
therefore not always able to match the specification precisely. The proof of
\eqref{e:GC-Snd} is by case analysis on $P$. We do not show the proof, rather we
demonstrate a proof later in this section which also synthesizes the definition
of $succ^{\sharp }$.

One advantage of the abstract interpretation framework is that it gives the
researcher the choice between four soundness properties, all of which are
equivalent and generated by $\alpha $ and $\gamma $:
\begin{align*}& \alpha (\mathord{\uparrow }succ(\gamma (P))) \subseteq  \mathord{\uparrow }succ^{\sharp }(P) \tag{GC-Snd/$\alpha \gamma $} \label{e:GC-Snd-ag}
\\& \mathord{\uparrow }succ(\gamma (P)) \subseteq  \gamma (\mathord{\uparrow }succ^{\sharp }(P)) \tag{GC-Snd/$\gamma \gamma $} \label{e:GC-Snd-gg}
\\& \alpha (\mathord{\uparrow }succ(N)) \subseteq  \mathord{\uparrow }succ^{\sharp }(\alpha (N)) \tag{GC-Snd/$\alpha \alpha $} \label{e:GC-Snd-aa}
\\& \mathord{\uparrow }succ(N) \subseteq  \gamma (\mathord{\uparrow }succ^{\sharp }(\alpha (N))) \tag{GC-Snd/$\gamma \alpha $} \label{e:GC-Snd-ga}
\end{align*}
Because each soundness property is equivalent (also a consequence of
\eqref{e:GC-Corr}), one can choose whichever variant is easiest to prove. The
soundness setup \eqref{e:GC-Snd} is the $\alpha \gamma $ rule, however any of the other rules can
also be used. For example, one could choose $\alpha \alpha $ or $\gamma \alpha $; in these cases the
proof considers four disjoint cases for $N$: $N$ is empty, $N$ contains only
even numbers, $N$ contains only odd numbers, and $N$ contains both even and odd
numbers.

\paragraph{Completeness}

The mappings $\alpha $ and $\gamma $ also synthesize an \emph{optimality} statement for
$\mathord{\uparrow }succ^{\sharp }$, a kind of completeness property, by stating that it
\emph{under}-approximates the optimal specification:
\begin{align*}& \alpha (\mathord{\uparrow }succ(\gamma (P))) \supseteq  \mathord{\uparrow }succ^{\sharp }(P) \end{align*}
Because the left-hand-side is an optimal specification, an abstract interpreter
will never be strictly more precise. Therefore, optimality is written
equivalently using an equality:
\begin{align*}& \alpha (\mathord{\uparrow }succ(\gamma (P))) = \mathord{\uparrow }succ^{\sharp }(P) \tag{GC-Opt} \label{e:GC-Cmp} \end{align*}
Not all analyzers are optimal, however optimality helps identify those which
approximate too much. Consider the analyzer $\mathord{\uparrow }succ^{\sharp \prime}$:
\begin{align*}& \mathord{\uparrow }succ^{\sharp \prime} : \wp (\mathbb{P}) \rightarrow  \wp (\mathbb{P}) && \mathord{\uparrow }succ^{\sharp \prime}(P) \coloneqq  \{ \texttt{\textsc{e}}\texttt{\textsc{v}}\texttt{\textsc{e}}\texttt{\textsc{n}},\texttt{\textsc{o}}\texttt{\textsc{d}}\texttt{\textsc{d}}\}  \end{align*}
This analyzer reports that $succ(n)$ could have any parity regardless of the
parity for $n$; it's the analyzer that always says ``I don't know''. This
analyzer is perfectly sound but non-optimal.

Just like soundness, four completeness statements are generated by $\alpha $ and $\gamma $,
however each of the statements are \emph{not} equivalent:
\begin{align*}& \alpha (\mathord{\uparrow }succ(\gamma (P))) = \mathord{\uparrow }succ^{\sharp }(P) \tag{GC-Cmp/$\alpha \gamma $} \label{e:GC-Cmp-ag}
\\& \mathord{\uparrow }succ(\gamma (P)) = \gamma (\mathord{\uparrow }succ^{\sharp }(P)) \tag{GC-Cmp/$\gamma \gamma $} \label{e:GC-Cmp-gg}
\\& \alpha (\mathord{\uparrow }succ(N)) = \mathord{\uparrow }succ^{\sharp }(\alpha (N)) \tag{GC-Cmp/$\alpha \alpha $} \label{e:GC-Cmp-aa}
\\& \mathord{\uparrow }succ(N) = \gamma (\mathord{\uparrow }succ^{\sharp }(\alpha (N))) \tag{GC-Cmp/$\gamma \alpha $} \label{e:GC-Cmp-ga}
\end{align*}
Abstract interpreters which satisfy the $\alpha \gamma $ variant are called
\emph{optimal} because they lose no more information than necessary, and those
which satisfy the $\gamma \alpha $ variant are called \emph{precise} because they lose no
information \emph{at all}. The abstract interpreter $succ^{\sharp }$ is optimal but not
precise, because $\gamma (\mathord{\uparrow }succ^{\sharp }(\alpha (\{ 1\} ))) \neq  \mathord{\uparrow }succ(\{ 1\} )$

To overcome mechanization issues with Galois connections, the state-of-the-art
is restricted to use $\gamma \gamma $ rules only for soundness \eqref{e:GC-Snd-gg} and
completeness \eqref{e:GC-Cmp-gg}. This is unfortunate for completeness properties
because each completeness variant is not equivalent. 

\paragraph{Calculational Derivation of Abstract Interpreters}

Rather than posit $\mathord{\uparrow }succ^{\sharp }$ and prove it correct directly, one can instead
derive its definition through a calculational process. The process begins with
the optimal specification on the left-hand-side of \eqref{e:GC-Cmp}, and reasons
equationally towards the definition of a function. In this way, $\mathord{\uparrow }succ^{\sharp }$ is not
postulated, rather it is derived by calculation, and the result is both sound
and optimal by construction.

The derivation is by case analysis on $P$ which has four cases: $\{ \} $, $\{ \texttt{\textsc{e}}\texttt{\textsc{v}}\texttt{\textsc{e}}\texttt{\textsc{n}}\} $,
$\{ \texttt{\textsc{o}}\texttt{\textsc{d}}\texttt{\textsc{d}}\} $ and $\{ \texttt{\textsc{e}}\texttt{\textsc{v}}\texttt{\textsc{e}}\texttt{\textsc{n}},\texttt{\textsc{o}}\texttt{\textsc{d}}\texttt{\textsc{d}}\} $; we show $P=\{ \texttt{\textsc{e}}\texttt{\textsc{v}}\texttt{\textsc{e}}\texttt{\textsc{n}}\} $:
\begin{align*}& \alpha (\mathord{\uparrow }succ(\gamma (\{ \texttt{\textsc{e}}\texttt{\textsc{v}}\texttt{\textsc{e}}\texttt{\textsc{n}}\} )))
\\& = \alpha (\mathord{\uparrow }succ(\{ n \mathrel{|} even(n)\} )) && \lbag \text{ defn. of $\gamma $ }\rbag 
\\& = \alpha (\{ succ(n) \mathrel{|} even(n)\} )  && \lbag \text{ defn. of $\mathord{\uparrow }succ$ }\rbag 
\\& = \alpha (\{ n \mathrel{|} odd(n)\} )         && \lbag \text{ defn. of $even/odd$ }\rbag 
\\& = \{ \texttt{\textsc{o}}\texttt{\textsc{d}}\texttt{\textsc{d}}\}                    && \lbag \text{ defn. of $\alpha $ }\rbag 
\\& \triangleq  \mathord{\uparrow }succ^{\sharp }(\{ \texttt{\textsc{e}}\texttt{\textsc{v}}\texttt{\textsc{e}}\texttt{\textsc{n}}\} )          && \lbag \text{ defining $\mathord{\uparrow }succ^{\sharp }$ }\rbag 
\end{align*}
The derivation of the other cases is analogous, and together they define the
implementation of $\mathord{\uparrow }succ^{\sharp }$.

Deriving analyzers by calculus is attractive because it is systematic, and
because it prevents the issue where an analyzer is postulated and discovered to
be unsound only after failing to complete its soundness proof. However, this
calculational style of abstract interpretation is not amenable to mechanized
verification with program extraction because $\alpha $ is often non-constructive, an
issue we describe later in this section.

\paragraph{Added Complexity}

The abstract interpretation approach requires a Galois connection up-front
which necessitates the introduction of powersets $\wp (\mathbb{N})$ and $\wp (\mathbb{P})$. This
results in powerset-lifted definitions and adds boilerplate set-theoretic
reasoning to the proofs.

This is in contrast to the direct approach which never mentions powersets of
parities. Not using powersets results in more understandable soundness
criteria, requires no boilerplate set-theoretic reasoning, and results in
fewer cases for the proof of soundness. This boilerplate becomes magnified in a
mechanized setting where all details must be spelled out to a proof assistant.
Furthermore, the simpler proof of \eqref{e:DA-SndS}---which was immediate from the
definition of $parity$---cannot be recovered within the abstract interpretation
framework, which shows one abandons simpler proof techniques in exchange for
the benefits of abstract interpretation.

\paragraph{Resistance to Mechanized Verification}

Despite the beauty and utility of Galois connections, advocates of the approach
have yet to reconcile their use with advances in mechanized reasoning:
\emph{every mechanized verification of an executable abstract interpreter
to-date has resisted the use of Galois connections, even when initially
designed to take advantage of the framework.}

The issue in mechanizing Galois connections amounts to a conflict between
supporting both classical set-theoretic reasoning and executable static
analyzers. Supporting executable static analyzers calls for constructive
mathematics, a problem for $\alpha $ functions because they are often
non-constructive, an observation first made by
Monniaux~\cite{local:monniaux:1998:ms-thesis}. To work around this limitation,
Pichardie~\cite{local:pichardie:2005:phd-thesis} advocates for designing abstract
interpreters which are merely inspired by Galois connections, but ultimately
avoiding their use in verification, which he terms the ``$\gamma $-only'' approach.
Successful verification projects such as Verasco adopt this ``$\gamma $-only''
approach~\cite{dvanhorn:Jourdan2015FormallyVerified, dvanhorn:Leroy2009Formal},
despite the use of Galois connections in designing the original Astr\'ee
analyzer~\cite{dvanhorn:Blanchet2003Static}.

To better understand the foundational issues with Galois connections and $\alpha $
functions, consider verifying the abstract interpretation approach to soundness
for our parity analyzer using a proof assistant built on constructive logic. In
this setting, the encoding of the Galois connection must support elements of
infinite powersets---like the set of all even numbers---as well as executable
abstract interpreters which manipulate elements of finite powersets---like
$\{ \texttt{\textsc{e}}\texttt{\textsc{v}}\texttt{\textsc{e}}\texttt{\textsc{n}},\texttt{\textsc{o}}\texttt{\textsc{d}}\texttt{\textsc{d}}\} $. To support representing infinite sets, the powerset $\wp (\mathbb{N})$ is
modelled constructively as a predicate $\mathbb{N} \rightarrow  prop$. To support defining
executable analyzers that manipulate sets of parities, the powerset $\wp (\mathbb{P})$ is
modelled as an enumeration of its inhabitants, which we call $\mathbb{P}^c$:
\begin{align*}& \mathbb{P}^c \coloneqq  \{ \texttt{\textsc{e}}\texttt{\textsc{v}}\texttt{\textsc{e}}\texttt{\textsc{n}},\texttt{\textsc{o}}\texttt{\textsc{d}}\texttt{\textsc{d}},\bot ,\top \}  \end{align*}
where $\bot $ and $\top $ represent $\{ \} $ and $\{ \texttt{\textsc{e}}\texttt{\textsc{v}}\texttt{\textsc{e}}\texttt{\textsc{n}},\texttt{\textsc{o}}\texttt{\textsc{d}}\texttt{\textsc{d}}\} $. This enables a
definition for $\mathord{\uparrow }succ^{\sharp } : \mathbb{P}^c \rightarrow  \mathbb{P}^c$ which can be extracted and executed. The
consequence of this design is a Galois connection between $\mathbb{N} \rightarrow  prop$ and $\mathbb{P}^c$;
the issue is now $\alpha $:
\begin{align*}& \alpha  : (\mathbb{N} \rightarrow  prop) \rightarrow  \mathbb{P}^c \end{align*}
This version of $\alpha $ cannot be defined constructively, as doing so requires
deciding predicates over $\phi  : \mathbb{N} \rightarrow  prop$. To define $\alpha $ one must perform case
analysis on predicates like $\exists n,\phi (n)\wedge even(n)$ to \emph{compute} an element of
$\mathbb{P}^c$, which is not possible for arbitrary $\phi $. However, $\gamma $ \emph{can} be
defined constructively:
\begin{align*}& \gamma  : \mathbb{P}^c \rightarrow  (\mathbb{N} \rightarrow  prop) \end{align*}
In general, any \emph{theorem} of soundness using Galois connections can be
rewritten to use only $\gamma $, making use of \eqref{e:GC-Corr}; this is the essence of
the ``$\gamma $-only'' approach, embodied by the soundness variant \eqref{e:GC-Snd-gg}.
However, this principle does not apply to all \emph{proofs} of soundness using
Galois connections, many of which mention $\alpha $ in practice. For example, the
$\gamma $-only setup does not support calculation in the style advocated by
Cousot~\cite{davdar:cousot:1999:calculational}. Furthermore, not all
\emph{completeness} theorems can be translated to $\gamma $-only style, such as
\eqref{e:GC-Cmp-ga} which is used to show an abstract interpreter is fully precise. 

\paragraph{Wrapping Up}

Abstract interpretation differs from the direct approach in which parts of the
design are postulated and which parts are derived. The direct approach requires
postulating the analyzer and definition of soundness. Using abstract
interpretation, a Galois connection between sets is postulated instead, and
definitions for soundness and completeness are synthesized from the Galois
connection. Also, abstract interpretation support deriving the definition of a
static analyzer directly from its proof of correctness.

The downside of abstract interpretation is that it requires lifting $succ$ and
$succ^{\sharp }$ into powersets, which results in boilerplate set-theoretic reasoning in
the proof of soundness. Finally, due to foundational issues, the abstract
interpretation framework is not amenable to mechanized verification while also
supporting program extraction using constructive logic.



\section{Constructive Galois Connections} \label{s:constructive-gcs} 

In this section we describe abstract interpretation with constructive Galois
connections---a parallel universe of Galois connections analogous to classical
ones. The framework enjoys all the benefits of abstract interpretation, but
like the direct approach avoids the pitfalls of added complexity and resistance
to mechanization.

We will describe the framework of constructive Galois connections between sets
$A$ and $B$. When instantiated to $\mathbb{N}$ and $\mathbb{P}$, the framework recovers exactly
the direct approach from Section~\ref{s:the-direct-approach}. We will also describe
constructive Galois connections in the absence of partial orders, or more
specifically, we will assume the discrete partial order: $x \sqsubseteq  y \Leftrightarrow  x = y$.
(Partial orders didn't appear in our demonstration of classical abstract
interpretation, but they are essential to the general theory.) We describe
generalizing to partial orders and recovering classical results from
constructive ones at the end of this section. The fully general theory of
constructive Galois connections is described in Section~\ref{s:metatheory} where it
is compared side-by-side to classical Galois connections.

\paragraph{Abstracting Sets}

A constructive Galois connection between sets $A$ and $B$ contains two
mappings: the first is called \emph{extraction}, notated $\eta $, and the second is
called \emph{interpretation}, notated $\mu $:
\begin{align*}& \eta  : A \rightarrow  B && \mu  : B \rightarrow  \wp (A) \end{align*}
$\eta $ and $\mu $ are analogous to classical Galois connection mappings $\alpha $ and $\gamma $.
In the parity analysis described in Section~\ref{s:the-direct-approach}, the
extraction function was $parity$ and the interpretation function was $\llbracket \textunderscore\rrbracket $.

Constructive Galois connection mappings $\eta $ and $\mu $ must form a correspondence
similar to \eqref{e:GC-Corr}:
\begin{align*}& x \in  \mu (y) \iff   \eta (x) = y \tag{CGC-Corr} \label{e:CGC-Corr} \end{align*}
The intuition behind the correspondence is the same as before: to compare an
element $x$ in $A$ to an element $y$ in $B$, it is equivalent to compare them
through either $\eta $ or $\mu $.

Like classical Galois connections, the correspondence between $\eta $ and $\mu $ is
stated equivalently through two composition laws. Extraction functions $\eta $
which form a constructive Galois connection are also a ``best abstraction'',
analogously to $\alpha $ in the classical setup:
\begin{align*}& sound : x \in  \mu (\eta (x))              \tag{CGC-Ext} \label{e:CGC-Ext}
\\& tight : x \in  \mu (y) \implies   \eta (x) = y     \tag{CGC-Red} \label{e:CGC-Red}
\end{align*}

\subparagraph{Aside} We use the term \emph{extraction function} and notation
$\eta $ from Nielson \emph{et} \emph{al}~\cite{dvanhorn:Neilson:1999} where $\eta $ is used
to simplify the definition of an abstraction function $\alpha $. We recover $\alpha $
functions from $\eta $ in a similar way. However, their treatment of $\eta $ is a
side-note to simplifying the definition of $\alpha $ and nothing more. We take this
simple idea much further to realize an entire theory of abstraction around
$\eta /\mu $ functions and their correspondences. In this ``lowered'' theory of $\eta /\mu $ we
describe soundness/optimality criteria and calculational derivations analogous
to that of $\alpha /\gamma $ while support mechanized verification, none of which is true
of Nielson \emph{et} \emph{al}'s use of $\eta $.

\paragraph{Induced Specifications}

Four equivalent soundness criteria are generated by $\eta $ and $\mu $ just like in
the classical framework. Each soundness statement uses $\eta $ and $\mu $ in a
different but equivalent way (assuming \eqref{e:CGC-Corr}). For a concrete $f : A \rightarrow 
A$ and abstract $f^{\sharp } : B \rightarrow  B$, $f^{\sharp }$ is sound \textit{i}\textit{f}\textit{f} any of the following properties
hold:
\begin{align*}& x \in  \mu (y) \wedge  y' = \eta (f(x)) \implies   y' = f^{\sharp }(y) \tag{CGC-Snd/$\eta \mu $} \label{e:CGC-Snd-em}
\\& x \in  \mu (y) \wedge  x' = f(x) \implies   x' \in  \mu (f^{\sharp }(y)) \tag{CGC-Snd/$\mu \mu $} \label{e:CGC-Snd-mm}
\\& y = \eta (f(x)) \implies   y = f^{\sharp }(\eta (x))           \tag{CGC-Snd/$\eta \eta $} \label{e:CGC-Snd-ee}
\\& x' = f(x) \implies   x' \in  \mu (f^{\sharp }(\eta (x)))         \tag{CGC-Snd/$\mu \eta $} \label{e:CGC-Snd-me}
\end{align*}
In the direct approach to verifying an example parity analysis described in
Section~\ref{s:the-direct-approach}, the first soundness property \eqref{e:DA-Snd} is
generated by the $\mu \mu $ variant, and the second soundness property \eqref{e:DA-SndS}
which enjoyed a simpler proof is generated by the $\eta \eta $ variant. We write these
soundness rules in a slightly strange way so we can write their completeness
analogs simply by replacing $\Rightarrow $ with $\Leftrightarrow $. The origin of these rules comes from
an adjunction framework, which we discuss in Section~\ref{s:metatheory}.

The mappings $\eta $ and $\mu $ also generate four completeness criteria which, like
classical Galois connections, are not equivalent:
\begin{align*}& x \in  \mu (y) \wedge  y' = \eta (f(x)) \iff   y' = f^{\sharp }(y) \tag{CGC-Cmp/$\eta \mu $} \label{e:CGC-Cmp-em}
\\& x \in  \mu (y) \wedge  x' = f(x) \iff   x' \in  \mu (f^{\sharp }(y)) \tag{CGC-Cmp/$\mu \mu $} \label{e:CGC-Cmp-mm}
\\& y = \eta (f(x)) \iff   y = f^{\sharp }(\eta (x))           \tag{CGC-Cmp/$\eta \eta $} \label{e:CGC-Cmp-ee}
\\& x' = f(x) \iff   x' \in  \mu (f^{\sharp }(\eta (x)))         \tag{CGC-Cmp/$\mu \eta $} \label{e:CGC-Cmp-me}
\end{align*}
Inspired by classical Galois connections, we call abstract interpreters $f^{\sharp }$
which satisfy the $\eta \mu $ variant \emph{optimal} and those which satisfy the $\mu \eta $
variant \emph{precise}.

The above soundness and completeness rules are stated for concrete and
abstraction \emph{functions} $f : A \rightarrow  A$ and $f^{\sharp } : B \rightarrow  B$. However, they
generalize easily to \emph{relations} $R : \wp (A\times A)$ and \emph{predicate
transformers} $F : \wp (A) \rightarrow  \wp (A)$ (\emph{i.e.} collecting semantics) through the
adjunction framework discussed in Section~\ref{s:metatheory}. The case studies in
Sections~\ref{s:case-cdgai} and~\ref{s:case-agt} describe abstract interpreters over
concrete relations and their soundness conditions.

\paragraph{Calculational Derivation of Abstract Interpreters}

The constructive Galois connection framework also supports deriving abstract
interpreters through calculation, analogously to the calculation we
demonstrated in Section~\ref{s:classical-ai}. To support calculational reasoning,
the four logical soundness criteria are rewritten into statements about
subsumption between powerset elements:
\begin{align*}& \{ \eta (f(x)) \mathrel{|} x \in  \mu (y)\}  \subseteq  \{ f^{\sharp }(y)\}  \tag{CGC-Snd/$\eta \mu $*} \label{e:CGC-Snd-meS}
\\& \{ f(x) \mathrel{|} x \in  \mu (y)\}  \subseteq  \mu (f^{\sharp }(y))   \tag{CGC-Snd/$\mu \mu $*} \label{e:CGC-Snd-mmS}
\\& \{ \eta (f(x))\}  \subseteq  \{ f^{\sharp }(\eta (x))\}          \tag{CGC-Snd/$\eta \eta $*} \label{e:CGC-Snd-eeS}
\\& \{ f(x)\}  \subseteq  \mu (f^{\sharp }(\eta (x)))           \tag{CGC-Snd/$\mu \eta $*} \label{e:CGC-Snd-emS}
\end{align*}
The completeness analog to the four rules replaces set subsumption with
equality. Using the $\eta \mu $* completeness rule, one calculates towards a
definition for $f^{\sharp }$ starting from the left-hand-side, which is the optimal
specification for abstract interpreters of $f$.

To demonstrate calculation using constructive Galois connections, we show the
derivation of $succ^{\sharp }$ from its induced specification, the result of which is
sound and optimal (because each step is $=$ in addition to $\subseteq $) by
construction; we show $p=\texttt{\textsc{e}}\texttt{\textsc{v}}\texttt{\textsc{e}}\texttt{\textsc{n}}$:
\begin{align*}& \{ parity(succ(n)) \mathrel{|} n \in  \llbracket \texttt{\textsc{e}}\texttt{\textsc{v}}\texttt{\textsc{e}}\texttt{\textsc{n}}\rrbracket \} 
\\& = \{ parity(succ(n)) \mathrel{|} even(n)\}    && \lbag \text{ defn. of $\llbracket \textunderscore\rrbracket $ }\rbag 
\\& = \{ flip(parity(n)) \mathrel{|} even(n)\}    && \lbag \text{ defn. of $parity$ }\rbag 
\\& = \{ flip(\texttt{\textsc{e}}\texttt{\textsc{v}}\texttt{\textsc{e}}\texttt{\textsc{n}})\}                   && \lbag \text{ Eq. ~\ref{e:DA-Corr} }\rbag 
\\& = \{ \texttt{\textsc{o}}\texttt{\textsc{d}}\texttt{\textsc{d}}\}                          && \lbag \text{ defn. of $flip$ }\rbag 
\\& \triangleq  \{ succ^{\sharp }(\texttt{\textsc{e}}\texttt{\textsc{v}}\texttt{\textsc{e}}\texttt{\textsc{n}})\}                  && \lbag \text{ defining $succ^{\sharp }$ }\rbag 
\end{align*}
We will show another perspective on this calculation later in this section,
where the derivation of $succ^{\sharp }$ is not only sound and optimal by construction,
but computable by construction as well.

\paragraph{Mechanized Verification}

In addition to the benefits of a general abstraction framework, constructive
Galois connections are amenable to mechanization in a way that classical Galois
connections are not. In our Agda library and case studies we mechanize
constructive Galois connections in full generality, as well as proofs that use
both mapping functions, such as calculational derivations.

As we discussed in Sections~\ref{s:the-direct-approach} and~\ref{s:classical-ai}, the
constructive encoding for infinite powersets $\wp (A)$ is $A \rightarrow  prop$.
This results in the following types for $\eta $ and $\mu $ when encoded
constructively:
\begin{align*}& \eta  : \mathbb{N} \rightarrow  \mathbb{P} && \mu  : \mathbb{P} \rightarrow  \mathbb{N} \rightarrow  prop \end{align*}
In constructive logic, the arrow type $\mathbb{N} \rightarrow  \mathbb{P}$ classifies computable functions,
and the arrow type $\mathbb{P} \rightarrow  \mathbb{N} \rightarrow  prop$ classifies undecidable relations.
\eqref{e:CGC-Corr} is then mechanized without issue:
\begin{align*}& \mu (p,n) \iff   \eta (n) = p \end{align*}
See the mechanization details in Section~\ref{s:the-direct-approach} for how $\eta $ and
$\mu $ are defined constructively for the example parity analysis.

\paragraph{Wrapping Up}

Constructive Galois connections are a general abstraction framework similar to
classical Galois connections. At the heart of the constructive Galois
connection framework is a correspondence \eqref{e:CGC-Corr} analogous to its
classical counterpart. From this correspondence, soundness and completeness
criteria are synthesized for abstract interpreters. Constructive Galois
connections also support calculational derivations of abstract interpreters
which and sound and optimal by construction. In addition to these benefits of a
general abstraction framework, constructive Galois connections are amenable to
mechanized verification. Both extraction ($\eta $) and interpretation ($\mu $) can be
mechanized effectively, as well as proofs of soundness, completeness, and
calculational derivations.

\subsection{Partial Orders and Monotonicity} \label{s:pos} 

The full theory of constructive Galois connections generalizes to posets
$\langle A,\sqsubseteq \rangle ^A$ and $\langle B,\sqsubseteq ^B\rangle $ by making the following changes:
\begin{itemize}\item Powersets must be downward-closed, that is for $X : \wp (A)$:
   \begin{align*}& x \in  X \wedge  x' \sqsubseteq  x \implies   x' \in  X \tag{PowerMon} \end{align*}
   Singleton sets $\{ x\} $ are reinterpreted to mean $\{ x' \mathrel{|} x' \sqsubseteq  x\} $. For
   mechanization, this means $\wp (A)$ is encoded as an \emph{antitonic} function,
   notated with a down-right arrow $A \shortsearrow  prop$, where the partial ordering on
   $prop$ is by implication.
\item Functions must be monotonic, that is for $f : A \rightarrow  A$:
   \begin{align*}& x \sqsubseteq  x' \implies   f(x) \sqsubseteq  f(x') \tag{FunMon} \end{align*}
   We notate monotonic functions $f : A \shortnearrow  A$. Monotonicity is required for
   mappings $\eta $ and $\mu $, and concrete and abstract interpreters $f$ and $f^{\sharp }$.
\item The constructive Galois connection correspondence is generalized to partial
   orders in place of equality, that is for $\eta $ and $\mu $:
   \begin{align*}& x \in  \mu (y) \iff   \eta (x) \sqsubseteq  y \tag{CGP-Corr} \end{align*}
   or alternatively, by generalizing the reductive property:
   \begin{align*}& x \in  \mu (y) \implies   \eta (x) \sqsubseteq  y \tag{CGP-Red} \end{align*}
\item Soundness criteria are also generalized to partial orders:
   \begin{align*}& x \in  \mu (y) \wedge  y' \sqsubseteq  \eta (f(x)) \implies   y' \sqsubseteq  f^{\sharp }(y) \tag{CGP-Snd/$\eta \mu $}
   \\& x \in  \mu (y) \wedge  x' \sqsubseteq  f(x) \implies   x' \in  \mu (f^{\sharp }(y)) \tag{CGP-Snd/$\mu \mu $}
   \\& y \sqsubseteq  \eta (f(x)) \implies   y \sqsubseteq  f^{\sharp }(\eta (x))           \tag{CGP-Snd/$\eta \eta $}
   \\& x' \sqsubseteq  f(x) \implies   x' \in  \mu (f^{\sharp }(\eta (x)))         \tag{CGP-Snd/$\mu \eta $}
   \end{align*}
   We were careful to write the equalities in Section~\ref{s:constructive-gcs} in
   the right order so this change is just swappping $=$ for $\sqsubseteq $. Completeness
   criteria are identical with $\Leftrightarrow $ in place of $\Rightarrow $.
\end{itemize}
To demonstrate when partial orders and monotonicity are necessary, consider
designing a parity analyzer for the $max$ operator:
\begin{align*}& max^{\sharp } : \mathbb{P} \times  \mathbb{P} \rightarrow  \mathbb{P}
\\& max^{\sharp }(\texttt{\textsc{e}}\texttt{\textsc{v}}\texttt{\textsc{e}}\texttt{\textsc{n}},\texttt{\textsc{e}}\texttt{\textsc{v}}\texttt{\textsc{e}}\texttt{\textsc{n}}) \coloneqq  \texttt{\textsc{e}}\texttt{\textsc{v}}\texttt{\textsc{e}}\texttt{\textsc{n}}  && max^{\sharp }(\texttt{\textsc{e}}\texttt{\textsc{v}}\texttt{\textsc{e}}\texttt{\textsc{n}},\texttt{\textsc{o}}\texttt{\textsc{d}}\texttt{\textsc{d}}) \coloneqq  \mathord{?}
\\& max^{\sharp }(\texttt{\textsc{o}}\texttt{\textsc{d}}\texttt{\textsc{d}},\texttt{\textsc{o}}\texttt{\textsc{d}}\texttt{\textsc{d}}) \coloneqq  \texttt{\textsc{o}}\texttt{\textsc{d}}\texttt{\textsc{d}}     && max^{\sharp }(\texttt{\textsc{o}}\texttt{\textsc{d}}\texttt{\textsc{d}},\texttt{\textsc{e}}\texttt{\textsc{v}}\texttt{\textsc{e}}\texttt{\textsc{n}}) \coloneqq  \mathord{?}
\end{align*}
The last two cases for $max^{\sharp }$ cannot be defined because the maximum of an even
and odd number could be either even or odd, and there is no representative for
``any number'' in $\mathbb{P}$. To remedy this, we add $\texttt{\textsc{a}}\texttt{\textsc{n}}\texttt{\textsc{y}}$ to the set of parities: $\mathbb{P}^+  \coloneqq 
\mathbb{P} \cup  \{ \texttt{\textsc{a}}\texttt{\textsc{n}}\texttt{\textsc{y}}\} $; the new element $\texttt{\textsc{a}}\texttt{\textsc{n}}\texttt{\textsc{y}}$ is interpreted: $\llbracket \texttt{\textsc{a}}\texttt{\textsc{n}}\texttt{\textsc{y}}\rrbracket  \coloneqq  \{ n \mathrel{|} n \in  \mathbb{N}\} $; the
partial order on $\mathbb{P}^+ $ becomes: $\texttt{\textsc{e}}\texttt{\textsc{v}}\texttt{\textsc{e}}\texttt{\textsc{n}},\texttt{\textsc{o}}\texttt{\textsc{d}}\texttt{\textsc{d}} \sqsubseteq  \texttt{\textsc{a}}\texttt{\textsc{n}}\texttt{\textsc{y}}$; and the correspondence
continues to hold using this partial order: $n \in  \llbracket p^+ \rrbracket  \iff   parity(n) \sqsubseteq  p^+ $.
$max^{\sharp }$ is then defined using the abstraction $\mathbb{P}^+ $ and proven sound and optimal
following the abstract interpretation paradigm.


\subsection{Relationship to Classical Galois Connections} \label{s:classical-rel} 

We clarify the relationship between constructive and classical Galois
connections in three ways:
\begin{itemize}\item Any constructive Galois connection can be lifted to obtain an equivalent
   classical Galois connection, and likewise for soundness and completeness
   proofs.
\item Any classical Galois connection which can be recovered by a constructive
   one contains no additional expressive power, rendering it an equivalent
   theory with added boilerplate reasoning.
\item Not all classical Galois connections can be recovered by constructive
   ones.
\end{itemize}
From these relationships we conclude that one benefits from using constructive
Galois connections whenever possible, classical Galois connections when no
constructive one exists, and both theories together as needed. We make these
claims precise in Section~\ref{s:metatheory}.

A classical Galois connection is recovered from a constructive one by the
following lifting:
\begin{align*}& \alpha  : \wp (A) \rightarrow  \wp (B) && \alpha (X) \coloneqq  \{ \eta (x) \mathrel{|} x \in  X\} 
\\& \gamma  : \wp (B) \rightarrow  \wp (A) && \gamma (Y) \coloneqq  \{ x \mathrel{|} y \in  Y \wedge  x \in  \mu (y)\} 
\end{align*}
When a classical Galois connection can be written in this form for some $\eta $ and
$\mu $, then one can use the simpler setting of abstract interpretation with
constructive Galois connections without any loss of generality. We also observe
that many classical Galois connections in practice can be written in this form,
and therefore can be mechanized effectively using constructive Galois
connections. The case studies in presented in Sections~\ref{s:case-cdgai}
and~\ref{s:case-agt} are two such cases, although the original authors of those
works did not initially write their classical Galois connections in this
explicitly lifted form.

An example of a classical Galois connection which is not recovered by lifting a
constructive Galois is the Independent Attributes (IA) abstraction, which
abstracts relations $R : \wp (A\times B)$ with their component-wise splitting $\langle R_l,R_r\rangle  :
\wp (A)\times \wp (B)$:
\begin{align*}& \alpha  : \wp (A\times B) \rightarrow  \wp (A)\times \wp (B)
\\& \alpha (R) \coloneqq  \langle \{ x \mathrel{|} \exists y. \langle x,y\rangle  \in  R\} ,\{ y \mathrel{|} \exists x. \langle x,y\rangle  \in  R\} \rangle 
\\& \gamma  : \wp (A)\times \wp (B) \rightarrow  \wp (A\times B)
\\& \gamma (R_l,R_r) \coloneqq  \{ \langle x,y\rangle  \mathrel{|} x \in  R_l, y \in  R_r\} 
\end{align*}
This Galois connection \emph{is} amenable to mechanized verification. In a
constructive setting, $\alpha $ and $\gamma $ are maps between $A\times B \rightarrow  prop$ and
$(A\rightarrow prop)\times (B\rightarrow prop)$, and can be defined directly using logical connectives $\exists $
and $\wedge $:
\begin{align*}& \alpha (R) \coloneqq  \langle \lambda (x).\exists (y).R(x,y),\lambda (y).\exists (x).R(x,y)\rangle 
\\& \gamma (R_l,R_r) \coloneqq  \lambda  (x,y). R_l(x) \wedge  R_r(y)
\end{align*}
IA can be mechanized effectively because the Galois connection consists of
mappings between specifications and the foundational issue of constructing
values from specifications does not appear. IA is not a constructive Galois
connection because there is no pure function $\mu $ underlying the abstraction
function $\alpha $.

Because constructive Galois connections can be lifted to classical ones, a
constructive Galois connection can interact directly with IA through its
lifting, even in a mechanized setting. However, once a constructive Galois
connection is lifted it loses its computational properties and cannot be
extracted and executed. In practice, IA is used to weaken ($\sqsubseteq $) an induced
optimal specification after which the calculated interpreter is shown to be
optimal ($=$) up-to-IA. IA never appears in the final calculated interpreter,
so not having a constructive Galois connection formulation poses no issue.


\subsection{The ``Specification Effect''} \label{s:spec-eff} 

The machinery of constructive Galois connections follow a \emph{monadic effect}
discipline, where the effect type is the classical powerset $\wp (\textunderscore)$; we call
this a \emph{specification effect}. First we will describe the monadic
structure of powersets $\wp (\textunderscore)$ and what we mean by ``specification effect''. Then
we will recast the theory of constructive Galois connections in this monadic
style, giving insights into why the theory supports mechanized verification,
and foreshadowing key fragments of the metatheory we develop in
Section~\ref{s:metatheory}.

The monadic structure of classical powersets is standard, and is analogous to
the nondeterminism monad familiar to Haskell programmers. However, the model
$\wp (A) = A \rightarrow  prop$ is the uncomputable nondeterminism monad and mirrors the use
of set-comprehensions on paper to describe uncomputable sets (specifications),
rather than the use of monad comprehensions in Haskell to describe computable
sets (constructed values).

We generalize $\wp (\textunderscore)$ to a \emph{monotonic} monad, similarly to how we
generalized powersets to posets in Section~\ref{s:pos}. This results in monotonic
versions of monad operators $ret$ and $bind$:
\begin{align*}& ret : A\! \shortnearrow \! \wp (A)            && bind : \wp (A)\times (A\! \shortnearrow \! \wp (B))\! \shortnearrow \! \wp (B)
\\& ret(x)\! \coloneqq \! \{ x' \mathrel{|} x'\! \sqsubseteq \! x\}     && bind(X,f)\! \coloneqq \! \{ y \mathrel{|} x\! \in \! X \wedge  y\! \in \! f(x)\} 
\end{align*}
We adopt Moggi's notation~\cite{davdar:Moggi:1989:Monads} for monadic extension where
$bind(X,f)$ is written $f^*(X)$, or just $f^*$ for $\lambda X.f^*(X)$.

We call the powerset type $\wp (A)$ a specification effect because it has monadic
structure, supports encoding arbitrary properties over values in $A$, and
cannot be ``escaped from'' in constructive logic, similar to the $IO$ monad in
Haskell. In classical mathematics, there is an isomorphism between singleton
powersets $\wp ^1 (A)$ and the set $A$. However, no such constructive mapping exists
for $\wp ^1 (A) \rightarrow  A$. Such a function would decide arbitrary predicates in $A \rightarrow 
prop$ to \emph{compute} the $A$ inside the singleton set. This observation,
that you can program inside $\wp (\textunderscore)$ monadically in constructive logic, but you
can't escape the monad, is why we call it a specification effect.

Given the monadic structure for powersets, and the intuition that they encode a
specification effect in constructive logic, we can recast the theory of
constructive Galois connections using monadic operators.
To do this we define a helper operator which injects ``pure'' functions into the
``effectful'' function space:
\begin{align*}& pure : (A\! \shortnearrow \! B)\! \shortnearrow \! (A\! \shortnearrow \! \wp (B)) && pure(f)(x) \coloneqq  ret(f(x))
\end{align*}
We then rewrite \eqref{e:CGC-Corr} using $ret$ and $pure$:
\begin{align*}& ret(x) \subseteq  \mu (y) \iff   pure(\eta )(x) \subseteq  ret(y) \tag{CGM-Corr} \end{align*}
and we rewrite the expansive and reductive variant of the correspondence using
$ret$, $bind$ (notated $f^*$) and $pure$:
\begin{align*}& ret(x) \subseteq  \mu ^*(pure(\eta )(x)) \tag{CGM-Exp}
\\& pure(\eta )^*(\mu (y)) \subseteq  ret(y) \tag{CGM-Red}
\end{align*}
The four soundness and completeness conditions can also be written in monadic
style; we show the $\eta \mu $ soundness property here:
\begin{align*}& pure(\eta )^*(pure(f)^*(\mu (y))) \subseteq  pure(f^{\sharp })(y) \tag{CGM-Snd} \label{e:CGM-Snd} \end{align*}
The left-hand-side of the ordering is the optimal specification for $f^{\sharp }$, just
like \eqref{e:CGC-Snd-em} but using monadic operators. The right-hand-side of the
ordering is $f^{\sharp }$ lifted to the monadic function space. The constructive
calculation of $succ^{\sharp }$ we showed earlier in this section is a calculation of
this form. The specification on the left has type $\wp (\mathbb{P})$, and it \emph{has
effects}, meaning it uses classical reasoning and can't be executed. The
abstract interpreter on the right also has type $\wp (\mathbb{P})$, but it \emph{has no
effects}, meaning it \emph{can} be extracted and executed. The calculated
abstract interpreter is thus not only sound and optimal by construction,
\emph{it is computable by construction}. 

Constructive Galois connections are empowering because they treat specification
like an effect, which optimal specifications \emph{ought to have}, and which
computable abstract interpreters \emph{ought not to have}. Using a monadic
effect discipline we support calculations which start with a specification
effect, and where the ``effect'' is eliminated through the process of
calculation. The monad laws are crucial in canceling uses of $ret$ with $bind$
to arrive at a final pure computation. For example, the first step in a
derivation for \eqref{e:CGM-Snd} can immediately simplify using monad laws to:
\begin{align*}& pure(\eta  \circ  f)^*(\mu (y)) \subseteq  pure(f^{\sharp })(y) \end{align*}



\section{Case Study: Calculational AI} \label{s:case-cdgai} 

In this section we apply constructive Galois connections to the
\emph{Calculational Design of a Generic Abstract Interpreter} from Cousot's
monograph~\cite{davdar:cousot:1999:calculational}. To our knowledge, \emph{we
achieve the first mechanically verified abstract interpreter derived by
calculus.} 

The key challenge in mechanizing the interpreter is supporting both abstraction
($\alpha $) and concretization ($\gamma $) mappings, which are required by the
calculational approach. Classical Galois connections do not support
mechanization of the abstraction mapping without the use of axioms, and the
required axioms block computation, preventing the extraction of verified
algorithms.

To verify Cousot's generic abstract interpreter we use constructive Galois
connections, which we describe in Section~\ref{s:constructive-gcs} and formalize in
Section~\ref{s:metatheory}. Using constructive Galois connections we encode
extraction ($\eta $) and interpretation ($\mu $) mappings as constructive analogs to
$\alpha $ and $\gamma $, calculate an abstract interpreter for an imperative programming
language which is sound and computable by construction, and recover the
original classical Galois connection results through a systematic lifting.

First we describe the setup for the analyzer: the abstract syntax, the concrete
semantics, and the constructive Galois connections involved. Following the
abstract interpretation paradigm with constructive Galois connections we design
abstract interpreters for denotation functions and semantics relations. We show
a fragment of our Agda mechanization which closely mirrors the pencil-and-paper
proof, as well as Cousot's original derivation.

\begin{figure} 
  \begin{align*}  i \in  &&\hspace{-1.5em}      \mathbb{Z} \coloneqq \ & \{ \ldots ,-1,0,1,\ldots \}                      && \text{integers}
  \\  b \in  &&\hspace{-1.5em}      \mathbb{B} \coloneqq \ & \{ true,false\}                      && \text{booleans}
  \\  x \in  &&\hspace{-1.5em}  \texttt{var} \Coloneqq \ & \ldots                                 && \text{variables}       
  \\  \oplus  \in  &&\hspace{-1.5em}  \texttt{aop} \Coloneqq \ & + \mathrel{|} - \mathrel{|} \times  \mathrel{|} /                    && \text{arithmetic op.}  
  \\  \olessthan  \in  &&\hspace{-1.5em}  \texttt{cmp} \Coloneqq \ & \mathbin{<} \mathrel{|} \mathbin{=}        && \text{comparison op.}  
  \\  \mathbin{\rotatebox[origin=c]{90}{$\olessthan$}} \in  &&\hspace{-1.5em}  \texttt{bop} \Coloneqq \ & \vee  \mathrel{|} \wedge                             && \text{boolean op.}     
  \\ ae \in  &&\hspace{-1.5em} \texttt{aexp} \Coloneqq \ & i \mathrel{|} x \mathrel{|} \texttt{rand} \mathrel{|} ae \oplus  ae         && \text{arithmetic exp.} 
  \\ be \in  &&\hspace{-1.5em} \texttt{bexp} \Coloneqq \ & b \mathrel{|} ae \olessthan  ae \mathrel{|} be \mathbin{\rotatebox[origin=c]{90}{$\olessthan$}} be            && \text{boolean exp.}    
  \\ ce \in  &&\hspace{-1.5em} \texttt{cexp} \Coloneqq \ & \texttt{skip} \mathrel{|} ce \mathbin{;} ce \mathrel{|} x \coloneqq  ae        &&                   
  \\      &&\hspace{-1.5em}        \mathrel{|}\ & \texttt{if}\ be\ \texttt{then}\ ce\ \texttt{else}\ ce &&                   
  \\      &&\hspace{-1.5em}        \mathrel{|}\ & \texttt{while}\ be\ \texttt{do}\ ce            && \text{command exp.}    
  \end{align*}
\vspace{-1em}\caption{Case Study: \texttt{WHILE} abstract syntax} \label{f:while-syntax} \vspace{-1em}\end{figure} 

\subsection{Concrete Semantics}

The \texttt{WHILE} language is an imperative programming language with arithmetic
expressions, variable assignment and while-loops. We show the syntax for this
language in Figure~\ref{f:while-syntax}. \texttt{WHILE} syntactically distinguished
arithmetic, boolean and command expressions. \texttt{rand} is an arithmetic expression
which can evaluate to any integer. Syntactic categories $\oplus $, $\olessthan $ and $\mathbin{\rotatebox[origin=c]{90}{$\olessthan$}}$ range
over arithmetic, comparison and boolean operators, and are introduced to
simplify the presentation. The \texttt{WHILE} language is taken from
Cousot's monograph~\cite{davdar:cousot:1999:calculational}.

The concrete semantics of \texttt{WHILE} is sketched without full definition in
Figure~\ref{f:while-concrete-sem}. Denotation functions $\llbracket \textunderscore\rrbracket ^a$, $\llbracket \textunderscore\rrbracket ^c$ and $\llbracket \textunderscore\rrbracket ^b$
give semantics to arithmetic, conditional and boolean operators. The semantics
of compound syntactic expressions are given operationally with relations
$\Downarrow ^a$, $\Downarrow ^b$ and $\mapsto ^c$. Relational semantics are given for arithmetic expressions
and commands due to the nondeterminism of $\texttt{rand}$ and nontermination of
$\texttt{while}$. These semantics serve as the starting point for designing an
abstract interpreter.

\subsection{Abstract Semantics with Constructive GCs}

Using abstract interpretation with constructive Galois connections, we design
an abstract semantics for \texttt{WHILE} in the following steps:
\begin{enumerate}\item An abstraction for each set $\mathbb{Z}$, $\mathbb{B}$ and $\texttt{env}$.
\item An abstraction for each denotation function $\llbracket \textunderscore\rrbracket ^a$, $\llbracket \textunderscore\rrbracket ^c$ and $\llbracket \textunderscore\rrbracket ^b$.
\item An abstraction for each semantics relation $\Downarrow ^a$, $\Downarrow ^b$ and $\mapsto ^c$.
\end{enumerate}
Each abstract set forms a constructive Galois connection with its concrete
counterpart. Soundness criteria is synthesized for abstract functions and
relations using constructive Galois connection mappings. Finally, we verify and
calculate abstract interpreters from these specifications which are sound and
computable by construction. We describe the details of this process only for
integers and environments (the sets $\mathbb{Z}$ and $\texttt{env}$), arithmetic operators (the
denotation function $\llbracket \textunderscore\rrbracket ^a$), and arithmetic expressions (the semantics relation
$\Downarrow ^a$). See the Agda development accompanying this paper for the full
mechanization of \texttt{WHILE}.

\begin{figure} 
  \begin{align*}    \rho  \in \ & \texttt{env} \coloneqq  \texttt{var} \rightharpoonup  \mathbb{Z}    & \varsigma       \in \ & \Sigma  \Coloneqq  \langle \rho ,ce\rangle             
  \\ \llbracket \textunderscore\rrbracket ^a \in \ & aop \rightarrow  \mathbb{Z} \times  \mathbb{Z} \rightharpoonup  \mathbb{Z}      & \textunderscore\vdash \textunderscore\Downarrow ^a\textunderscore \in \ & \wp (\texttt{env} \times  \texttt{aexp} \times  \mathbb{Z}) 
  \\ \llbracket \textunderscore\rrbracket ^c \in \ & cmp \rightarrow  \mathbb{Z} \times  \mathbb{Z} \rightarrow  \mathbb{B}      & \textunderscore\vdash \textunderscore\Downarrow ^b\textunderscore \in \ & \wp (\texttt{env} \times  \texttt{bexp} \times  \mathbb{B}) 
  \\ \llbracket \textunderscore\rrbracket ^b \in \ & bop \rightarrow  \mathbb{B} \times  \mathbb{B} \rightarrow  \mathbb{B}      & \textunderscore\mapsto ^c\textunderscore    \in \ & \wp (\Sigma  \times  \Sigma )              
  \end{align*} \vspace{-2em}
  \begin{mathpar} \inferrule*[right=ARand]
     {\ }
     {\rho  \vdash  \texttt{rand} \Downarrow ^a i}
  
 \inferrule*[right=AOp]
     {\rho  \vdash  ae_1  \Downarrow ^a i_1  \\ \rho  \vdash  ae_2  \Downarrow ^a i_2 }
     {\rho  \vdash  ae_1  \oplus  ae_2  \Downarrow ^a \llbracket \oplus \rrbracket ^a(i_1 ,i_2 )}
  
 \inferrule*[right=CAssign]
     {\rho  \vdash  ae \Downarrow ^a i}
     {\langle \rho ,x \coloneqq  ae\rangle  \mapsto ^c \langle \rho [x\leftarrow i],\texttt{skip}\rangle }
  
 \inferrule*[right=CWhile-T]
     {\rho  \vdash  be \Downarrow ^b true}
     {\langle \rho ,\texttt{while}\ be\ \texttt{do}\ ce\rangle  \mapsto ^c \langle \rho ,ce \mathbin{;} \texttt{while}\ be\ \texttt{do}\ ce\rangle }
  
 \inferrule*[right=CWhile-F]
     {\rho  \vdash  be \Downarrow ^b false}
     {\langle \rho ,\texttt{while}\ be\ \texttt{do}\ ce\rangle  \mapsto ^c \langle \rho ,\texttt{skip}\rangle }
  \end{mathpar}
\vspace{-1em}\caption{Case Study: \texttt{WHILE} concrete semantics} \label{f:while-concrete-sem} \vspace{-1em}\end{figure} 

\paragraph{Abstracting Integers}

We design a simple sign abstraction for integers, although more powerful
abstractions are certainly possible~\cite{dvanhorn:Mine2006Octagon}. The final
abstract interpreter for \texttt{WHILE} is parameterized by any abstraction for
integers, meaning another abstraction can be plugged in without added proof
effort. 

The sign abstraction begins with three representative elements: \texttt{neg}, \texttt{zer}
and \texttt{pos}, representing negative integers, the integer $0$, and positive
integers. To support representing integers which could be negative or $0$,
negative or positive, or $0$ or positive, etc. we design a set which is
complete w.r.t these logical disjunctions:
\begin{align*}& i^{\sharp } \in  \mathbb{Z}^{\sharp } \coloneqq  \{ \texttt{none},\texttt{neg},\texttt{zer},\texttt{pos},\texttt{negz},\texttt{nzer},\texttt{posz},\texttt{any}\}  \end{align*}
$\mathbb{Z}^{\sharp }$ is given meaning through an interpretation function $\mu ^z$, the analog of a
$\gamma $ from the classical Galois connection framework:
\begin{align*}& \mu ^z : \mathbb{Z}^{\sharp } \shortnearrow  \wp (\mathbb{Z})           &&
\\& \mu ^z(\texttt{none}) \coloneqq  \{ \}           && \mu ^z(\texttt{negz}) \coloneqq  \{ i \mathrel{|} i \leq  0\} 
\\& \mu ^z(\texttt{neg}) \coloneqq  \{ i \mathrel{|} i < 0\}   && \mu ^z(\texttt{nzer}) \coloneqq  \{ i \mathrel{|} i \neq  0\} 
\\& \mu ^z(\texttt{zer}) \coloneqq  \{ 0\}           && \mu ^z(\texttt{posz}) \coloneqq  \{ i \mathrel{|} i \geq  0\} 
\\& \mu ^z(\texttt{pos}) \coloneqq  \{ i \mathrel{|} i > 0\}   && \mu ^z(\texttt{any}) \coloneqq  \{ i \mathrel{|} i \in  \mathbb{Z}\} 
\end{align*}
The partial ordering on abstract integers coincides with subset ordering
through $\mu ^z$, that is $i^{\sharp }_1  \sqsubseteq ^z i^{\sharp }_2  \iff   \mu ^z(i^{\sharp }_1 ) \subseteq  \mu ^z(i^{\sharp }_2 )$:
\begin{align*}& \texttt{neg}\sqsubseteq ^z\texttt{negz},\texttt{nzer} && \texttt{pos}\sqsubseteq ^z\texttt{nzer},\texttt{posz}
\\& \texttt{zer}\sqsubseteq ^z\texttt{negz},\texttt{posz} && \texttt{none} \sqsubseteq ^z i^{\sharp } \sqsubseteq ^z i^{\sharp } \sqsubseteq ^z \texttt{any} 
\end{align*}
To be a constructive Galois connection, $\mu ^z$ forms a correspondence with a best
abstraction function $\eta ^z$:
\begin{align*}& \eta ^z : \mathbb{Z} \rightarrow  \mathbb{Z}^{\sharp }
&& \eta ^z(n) \coloneqq  \begin{cases}\texttt{neg} & \textit{i}\textit{f}\; i < 0
                     \\\texttt{zer} & \textit{i}\textit{f}\; i = 0
                     \\\texttt{pos} & \textit{i}\textit{f}\; i > 0
          \end{cases}
\end{align*}
and we prove the constructive Galois connection correspondence:
\begin{align*}& i \in  \mu ^z(i^{\sharp }) \iff   \eta ^z(i) \sqsubseteq ^z i^{\sharp } \end{align*}

\subparagraph{The Classical Design}
To contrast with Cousot's original design using classical abstract
interpretation, the key difference is the abstraction function. The abstraction
function using classical Galois connections is recovered through a lifting of
our $\eta ^z$:
\begin{align*}& \alpha ^z : \wp (\mathbb{Z}) \shortnearrow  \mathbb{Z}^{\sharp } && \alpha ^z(I) \coloneqq  \bigsqcup _{i\in I}\eta ^z(i)
\end{align*}
Abstraction functions of this form---$\wp (B) \shortnearrow  A$, for some concrete set $A$ and
abstract set $B$---are representative of most Galois connections used in the
literature for static analyzers. However, these abstraction functions are
precisely the part of classical Galois connections which inhibit mechanized
verification. The extraction function $\eta ^z$ does not manipulate powersets, does
not inhibit mechanized verification, and recovers the original non-constructive
$\alpha ^z$ through this standard lifting.

\paragraph{Abstracting Environments}

An abstract environment maps variables to abstract integers rather than
concrete integers.
\begin{align*}& \rho ^{\sharp } \in  \texttt{env}^{\sharp } \coloneqq  \texttt{var} \rightarrow  \mathbb{Z}^{\sharp } \end{align*}
$\texttt{env}^{\sharp }$ is given meaning through an interpretation function $\mu ^r$:
\begin{align*}& \mu ^r \in  \texttt{env}^{\sharp }\! \shortnearrow \! \wp (\texttt{env}) && \mu ^r(\rho ^{\sharp }) \coloneqq  \{  \rho  \mathrel{|} \forall x.\rho (x) \in  \mu ^z(\rho ^{\sharp }(x)) \} 
\end{align*}
An abstract environment represents concrete environments that agree pointwise
with some represented integer in the codomain.

The order on abstract environments is the standard pointwise ordering and obeys
$\rho ^{\sharp }_1  \sqsubseteq ^r \rho ^{\sharp }_2  \iff   \mu ^r(\rho ^{\sharp }_1 ) \subseteq  \mu ^r(\rho ^{\sharp }_2 )$:
\begin{align*}& \rho ^{\sharp }_1  \sqsubseteq ^r \rho ^{\sharp }_2  \iff   (\forall x.\rho ^{\sharp }_1 (x) \sqsubseteq ^z \rho ^{\sharp }_2 (x)) \end{align*}
To form a constructive Galois connection, $\mu ^r$ forms a correspondence with a
best abstraction function $\eta ^r$:
\begin{align*}& \eta ^r \in  \texttt{env} \rightarrow  \texttt{env}^{\sharp } && \eta ^r(\rho ) \coloneqq  \lambda x. \eta ^z(\rho (x))
\end{align*}
and we prove the constructive Galois connection correspondence:
\begin{align*}& \rho  \in  \mu ^r(\rho ^{\sharp }) \iff   \eta ^r(\rho ) \sqsubseteq ^r \rho ^{\sharp } \end{align*}

\subparagraph{The Classical Design}
To contrast with Cousot's original design using classical abstract
interpretation, the key difference is again the abstraction function. The
abstraction function using classical Galois connections is:
\begin{align*}& \alpha ^r : \wp (\texttt{env})\! \shortnearrow \! \texttt{env}^{\sharp } && \alpha ^r(R) \coloneqq  \lambda x.\alpha ^z(\{ \rho (x) \mathrel{|} \rho  \in  R\} )
\end{align*}
which is also not amenable to mechanized verification.

\paragraph{Abstracting Functions}

After designing constructive Galois connections for $\mathbb{Z}$ and \texttt{env} we define
what it means for $\llbracket \textunderscore\rrbracket ^{\sharp a}$, some abstract denotation for arithmetic operators,
to be a sound abstraction of $\llbracket \textunderscore\rrbracket ^a$, its concrete counterpart. This is done
through a specification induced by mappings $\eta $ and $\mu $, analogously to how
specifications are induced using classical Galois connections.

The specification which encodes soundness and optimality for $\llbracket \textunderscore\rrbracket ^{\sharp a}$ is
generated using the constructive Galois connection for $\mathbb{Z}$:
\begin{align*}&   \langle \!i_1 \!,\!i_2 \!\rangle \! \in \! \mu \!^z\!(\!i^{\sharp }_1 \!,\!i^{\sharp }_2 \!) 
  \wedge  \langle \!i^{\sharp \prime}_1 \!,\!i^{\sharp \prime}_2 \!\rangle \! \sqsubseteq \! \eta \!^z\!(\llbracket ae\rrbracket ^a\!(\!i_1 \!,\!i_2 \!)) 
  \Leftrightarrow  \langle \!i^{\sharp \prime}_1 \!,\!i^{\sharp \prime}_2 \!\rangle \! \sqsubseteq \! \llbracket ae\rrbracket ^{\sharp a}\!(\!i^{\sharp }_1 \!,\!i^{\sharp }_2 \!) 
\end{align*}
(See \eqref{e:CGC-Cmp-em} in Section~\ref{s:constructive-gcs} for the origin of this
equation.) For $\llbracket \textunderscore\rrbracket ^{\sharp a}$, we postulate its definition and verify its correctness
post-facto using the above property, although we omit the proof details here.
The definition of $\llbracket \textunderscore\rrbracket ^{\sharp a}$ is standard, and returns \texttt{none} in the case of
division by zero. We show only the definition of $+$ here:
\begin{align*}& \llbracket \textunderscore\rrbracket ^{\sharp a} : \texttt{aexp} \rightarrow  \mathbb{Z}^{\sharp } \times  \mathbb{Z}^{\sharp } \shortnearrow  \mathbb{Z}^{\sharp } 
\\& \llbracket +\rrbracket ^{\sharp a}(i^{\sharp }_1 ,i^{\sharp }_2 ) \coloneqq  \bigsqcup \begin{cases} \texttt{pos} & \textit{i}\textit{f}\; \texttt{pos} \sqsubseteq ^z i^{\sharp }_1  \vee  \texttt{pos} \sqsubseteq ^z i^{\sharp }_2 
                               \\ \texttt{neg} & \textit{i}\textit{f}\; \texttt{neg} \sqsubseteq ^z i^{\sharp }_1  \vee  \texttt{neg} \sqsubseteq ^z i^{\sharp }_2 
                               \\ \texttt{zer} & \textit{i}\textit{f}\; \texttt{zer} \sqsubseteq ^z i^{\sharp }_1  \wedge  \texttt{zer} \sqsubseteq ^z i^{\sharp }_2 
                               \\ \texttt{zer} & \textit{i}\textit{f}\; \texttt{pos} \sqsubseteq ^z i^{\sharp }_1  \wedge  \texttt{neg} \sqsubseteq ^z i^{\sharp }_2 
                               \\ \texttt{zer} & \textit{i}\textit{f}\; \texttt{neg} \sqsubseteq ^z i^{\sharp }_1  \wedge  \texttt{pos} \sqsubseteq ^z i^{\sharp }_2 
                    \end{cases}
\end{align*}

\subparagraph{The Classical Design}
To contrast with Cousot's original design using classical abstract
interpretation, the key difference is that we avoid powerset liftings
all-together. Using classical Galois connections, the concrete denotation
function must be lifted to powersets:
\begin{align*}& \llbracket \textunderscore\rrbracket ^a_{\wp } \in  \texttt{aexp} \rightarrow  \wp (\mathbb{Z} \times  \mathbb{Z}) \rightarrow  \wp (\mathbb{Z})
\\& \llbracket ae\rrbracket ^a_{\wp }(II) \coloneqq  \{  \llbracket ae\rrbracket ^a(i_1 ,i_2 ) \mathrel{|} \langle i_1 ,i_2 \rangle  \in  II \} 
\end{align*}
and then $\llbracket \textunderscore\rrbracket ^{\sharp a}$ is proven correct w.r.t. this lifting using $\alpha ^z$ and $\gamma ^z$:
\begin{align*}& \alpha ^z(\llbracket ae\rrbracket ^a_{\wp }(\gamma ^z(i^{\sharp }_1 ,i^{\sharp }_2 ))) = \llbracket ae\rrbracket ^{\sharp a}(i^{\sharp }_1 ,i^{\sharp }_2 ) \end{align*}
This property cannot be mechanized without axioms because $\alpha ^z$ is
non-constructive. Furthermore, the proof involves additional powerset
boilerplate reasoning, which is not present in our mechanization of correctness
for $\llbracket \textunderscore\rrbracket ^{\sharp a}$ using constructive Galois connections. The state-of-the art
approach of ``$\gamma $-only'' verification would instead mechanize the $\gamma \gamma $ variant of
correctness:
\begin{align*}& \llbracket ae\rrbracket ^a_{\wp }(\gamma ^z(i^{\sharp }_1 ,i^{\sharp }_2 )) = \gamma ^z(\llbracket ae\rrbracket ^{\sharp a}(i^{\sharp }_1 ,i^{\sharp }_2 )) \end{align*}
which is similar to our $\mu \mu $ rule:
\begin{align*}&   \langle \!i_1 \!,i_2 \!\rangle \! \in \! \mu \!^z\!(\!i^{\sharp }_1 \!,\!i^{\sharp }_2 \!) 
  \wedge  \langle \!i'_1 \!,i'_2 \!\rangle \! =\! \llbracket ae\rrbracket ^a\!(\!i_1 \!,\!i_2 \!) 
  \Leftrightarrow  \langle \!i'_1 \!,i'_2 \!\rangle \! \in \! \mu \!^z\!(\llbracket ae\rrbracket ^{\sharp a}\!(\!i^{\sharp }_1 \!,\!i^{\sharp }_2 \!)) 
\end{align*}
The benefit of our approach is that soundness and completeness properties which
also mention extraction ($\eta $) can also be mechanized, like calculating abstract
interpreters from their specification.

\paragraph{Abstracting Relations}

The verification of an abstract interpreter for relations is similar to the
design for functions: induce a specification using the constructive Galois
connection, and prove correctness w.r.t. the induced spec. The relations we
abstract are $\Downarrow ^a$, $\Downarrow ^b$ and $\mapsto ^c$, and we call their abstract interpreters $\mathcal{A}^{\sharp }$,
$\mathcal{B}^{\sharp }$ and $\mathcal{C}^{\sharp }$. Rather than postulate the definitions of the abstract
interpreters, we calculate them from their specifications, the results of which
are sound and computable by construction. The arithmetic and boolean abstract
interpreters are functions from abstract environments to abstract integers, and
the abstract interpreter for commands computes the next abstract transition
states of execution. We only present select calculations for $\mathcal{A}^{\sharp }$; see our
accompanying Agda development for each calculation in mechanized form. $\mathcal{A}^{\sharp }$ has
type:
\begin{align*}& \mathcal{A}^{\sharp }[\textunderscore] : \texttt{aexp} \rightarrow  \texttt{env}^{\sharp } \shortnearrow  \mathbb{Z}^{\sharp } \end{align*}
To induce a spec for $\mathcal{A}^{\sharp }$, we first revisit the concrete semantics relation as
a powerset-valued function, which we call $\mathcal{A}$:
\begin{align*}& \mathcal{A}[\textunderscore] : \texttt{aexp} \rightarrow  \texttt{env} \rightarrow  \wp (\mathbb{Z}) && \mathcal{A}[ae](\rho ) \coloneqq  \{ i \mathrel{|} \rho  \vdash  ae \Downarrow ^a i\} 
\end{align*}
The induced spec for $\mathcal{A}^{\sharp }$ is generated with the monadic bind operator, which we
notate using Moggi's star notation $\textunderscore^*$:
\begin{align*}& pure(\eta ^z)^*(\mathcal{A}[ae]^*(\mu ^r(\rho ^{\sharp }))) \subseteq  pure(\mathcal{A}^{\sharp }[ae])(\rho ^{\sharp }) \end{align*}
which unfolds to:
\begin{align*}& \{ \eta ^z(i) \mathrel{|} \rho  \in  \mu ^r(\rho ^{\sharp }) \wedge  \rho  \vdash  ae \Downarrow ^a i\}  \subseteq  \{ \mathcal{A}^{\sharp }[ae](\rho ^{\sharp })\}  \end{align*}
To calculate $\mathcal{A}^{\sharp }$ we reason equationally from the spec on the left towards the
singleton set on the right, and declare the result the definition of $\mathcal{A}^{\sharp }$. We
do this by case analysis on $ae$; we show the cases for $ae=\texttt{rand}$ and $ae=x$
in Figure~\ref{f:calc}. Each calculation can also be written in monadic form, which
is the style we mechanize; we repeat the variable case in monadic form in the
figure.

\begin{figure} 
  \begin{align*}& \textbf{Case $ae=\texttt{rand}$:}
  \\& \{ \eta ^z(i) \mathrel{|} \rho  \in  \mu ^r(\rho ^{\sharp }) \wedge  \rho  \vdash  \texttt{rand} \Downarrow ^a i\}  \hspace{-4em}
  \\& = \{ \eta ^z(i) \mathrel{|} \rho  \in  \mu ^r(\rho ^{\sharp }) \wedge  i \in  \mathbb{Z}\}              && \lbag \text{ defn. of $\rho  \vdash  \texttt{rand} \Downarrow ^a i$ }\rbag 
  \\& \subseteq  \{ \eta ^z(i) \mathrel{|} i \in  \mathbb{Z}\}                           && \lbag \text{ $\varnothing $ when $\mu ^r(\rho ^{\sharp })=\varnothing $ }\rbag 
  \\& \subseteq  \{ \texttt{any}\}                                   && \lbag \text{ $\{ \texttt{any}\} $ mon. w.r.t. $\sqsubseteq ^z$ }\rbag 
  \\& \triangleq  \{ \mathcal{A}^{\sharp }[\texttt{rand}](\rho ^{\sharp })\}                          && \lbag \text{ defining $\mathcal{A}^{\sharp }[\texttt{rand}]$ }\rbag 
  \\& \textbf{Case $ae=x$:}
  \\& \{ \eta ^z(i) \mathrel{|} \rho  \in  \mu ^r(\rho ^{\sharp }) \wedge  \rho  \vdash  x \Downarrow ^a i\} 
  \\& = \{ \eta ^z(\rho (x)) \mathrel{|} \rho  \in  \mu ^r(\rho ^{\sharp })\}                   && \lbag \text{ defn. of $\rho  \vdash  x \Downarrow ^a i$ }\rbag 
  \\& = \{ \eta ^z(i) \mathrel{|} i \in  \mu ^z(\rho ^{\sharp }(x))\}                   && \lbag \text{ defn. of $\mu ^r(\rho ^{\sharp })$ }\rbag 
  \\& \subseteq  \{ \rho ^{\sharp }(x)\}                                   && \lbag \text{ Eq. ~\ref{e:CGC-Red} }\rbag 
  \\& \triangleq  \{ \mathcal{A}^{\sharp }[x](\rho ^{\sharp })\}                               && \lbag \text{ defining $\mathcal{A}^{\sharp }[x]$ }\rbag 
  \\& \textbf{Case $ae=x$ (Monadic):}
  \\& pure(\eta ^z)^*(\mathcal{A}[x]^*(\mu ^r(\rho ^{\sharp })))
  \\& = pure(\lambda \rho .\eta ^z(\rho (x)))^*(\mu ^r(\rho ^{\sharp }))               && \lbag \text{ defn. of $\mathcal{A}[x]$ }\rbag 
  \\& = pure(\eta ^z)^*(\mu ^{z*}(\rho ^{\sharp }(x)))                  && \lbag \text{ defn. of $\mu ^r(\rho ^{\sharp })$ }\rbag 
  \\& \subseteq  ret(\rho ^{\sharp }(x))                               && \lbag \text{ Eq. ~\ref{e:CGC-Red} }\rbag 
  \\& \triangleq  pure(\mathcal{A}^{\sharp }[x])(\rho ^{\sharp })                          && \lbag \text{ defining $\mathcal{A}^{\sharp }[x]$ }\rbag 
  \end{align*}
\vspace{-1em}\caption{Constructive GC calculations on paper} \label{f:calc} \vspace{-1em}\end{figure} 

\paragraph{Mechanized Calculation}

Our Agda calculation of $\mathcal{A}^{\sharp }$ strongly resembles the on-paper monadic one. We
show the Agda proof code for abstract variable references in
Figure~\ref{f:calc-agda}. The first line is the top level definition site for the
derivation of $\mathcal{A}^{\sharp }$ for the \texttt{Var} case. The \texttt{proof-mode} term is part of our
``proof-mode'' library which gives support for calculational reasoning in the
form of Agda proof combinators with mixfix syntax. Statements surrounded by
double square brackets $[[e]]$ restate the current proof state, which Agda will
check is correct. Reasoning steps are employed through $\lbag \text{e}\rbag $ terms, which
transform the proof state from the previous form to the next. The term
$[\texttt{focus-right}\; [\cdotp ]\; of\; e]$ focuses the goal to the right of the outermost
application, scoped between \texttt{begin} and \texttt{end}.

Using constructive Galois connections, our mechanized calculation closely
follows Cousot's classical one, uses both $\eta $ and $\mu $ mappings, and results in
a verified, executable static analyzer. Such a result is not possible using
classical Galois connections, due to the inability to encode $\alpha $ functions
constructively.

We complete the full calculation of Cousot's generic abstract interpreter for
\texttt{WHILE} in Agda as supplemental material to this paper, where the resulting
interpreter is both sound and computable by construction. We also provide our
``proof-mode'' library which supports general calculational reasoning with
posets.

\subparagraph{The Classical Design}
Classically, one first designs a powerset lifting of the concrete semantics,
called a \emph{collecting semantics}:
\begin{align*}& \mathcal{A}_{\wp }[\textunderscore] : \texttt{aexp} \rightarrow  \wp (\texttt{env}) \shortnearrow  \wp (\mathbb{Z})
\\& \mathcal{A}_{\wp }[ae](R) \coloneqq  \{  i \mathrel{|} \rho  \in  R \wedge  \rho  \vdash  ae \Downarrow ^a \} 
\end{align*}
The classical soundness specification for $\mathcal{A}^{\sharp }[ae](\rho ^{\sharp })$ is then:
\begin{align*}& \alpha ^z(\mathcal{A}_{\wp }[ae](\gamma ^r(\rho ^{\sharp }))) \sqsubseteq  \mathcal{A}^{\sharp }[ae](\rho ^{\sharp }) \end{align*}
However, as usual, the abstraction $\alpha ^z$ cannot be mechanized effectively,
preventing a mechanized derivation of $\mathcal{A}^{\sharp }$ by calculus.


\section{Case Study: Gradual Type Systems} \label{s:case-agt} 

Recent work in metatheory for gradual type systems
by \citet{local:garcia:2016:agt} shows how a Galois connection discipline can
guide the design of gradual typing systems. Starting with a Galois connection
between precise and gradual types, both the static and dynamic semantics of the
gradual language are derived systematically. This technique is called
Abstracting Gradual Typing (AGT).

The design presented by Garcia et al is to begin with a precise type system,
like the simply typed lambda calculus, and add a new type $\mathord{?}$ which functions
as the $\top $ element in the lattice of type precision. The precise typing rules
are presented with meta-operators $<:$ for subtyping and $\overset{..}{\vee}$ for the join
operator in the subtyping lattice. The gradual type system is then written
using abstract variants $<:^{\sharp }$ and $\overset{..}{\vee}^{\sharp }$ which are proven correct w.r.t.
specifications induced by the Galois connection. 

\begin{figure} 
  \begin{align*}& \texttt{-- Agda Calculation of Case $ae=x$: }
  \\& \alpha [\mathcal{A}]\; (\texttt{Var}\; x)\; \rho \sharp\; =\; \texttt{[proof-mode]}
  \\& \hspace{1em}\texttt{do}\; [[\; (\texttt{pure}\; \cdotp \; \eta ^z)\; *\; \cdotp \; (\mathcal{A}[\; \texttt{Var}\; x\; ]\; *\; \cdotp \; (\mu ^r\; \cdotp \; \rho \sharp))\; ]]
  \\& \hspace{1em}\;\ \!\centerdot\;    [\texttt{focus-right}\; [\cdotp ]\; of\; (\texttt{pure}\; \cdotp \; \eta ^z)\; *\; ]\; \texttt{begin}
  \\& \hspace{1em}\hspace{1em}\texttt{do}\; [[\; \mathcal{A}[\; \texttt{Var}\; x\; ]\; *\; \cdotp \; (\mu ^r\; \cdotp \; \rho \sharp)\; ]]
  \\& \hspace{1em}\hspace{1em}\;\ \!\centerdot\;    \lbag\; \mathcal{A}[\texttt{Var}]/\mathord{\equiv }\; \rbag
  \\& \hspace{1em}\hspace{1em}\;\ \!\centerdot\;    [[\; (\texttt{pure}\; \cdotp \; \texttt{lookup}[\; x\; ])\; *\; \cdotp \; (\mu ^r\; \cdotp \; \rho \sharp)\; ]]
  \\& \hspace{1em}\hspace{1em}\;\ \!\centerdot\;    \lbag\; \texttt{lookup}/\mu ^r/\mathord{\equiv }\; \rbag
  \\& \hspace{1em}\hspace{1em}\;\ \!\centerdot\;    [[\; \mu ^z\; *\; \cdotp \; (\texttt{pure}\; \cdotp \; \texttt{lookup}\sharp[\; x\; ]\; \cdotp \; \rho \sharp)\; ]]
  \\& \hspace{1em}\hspace{1em}\texttt{end}
  \\& \hspace{1em}\;\ \!\centerdot\;    [[\; (\texttt{pure}\; \cdotp \; \eta ^z)\; *\; \cdotp \; (\mu ^z\; *\; \cdotp \; (\texttt{pure}\; \cdotp \; \texttt{lookup}\sharp[\; x\; ]\; \cdotp \; \rho \sharp))\; ]]
  \\& \hspace{1em}\;\ \!\centerdot\;    \lbag\; \texttt{reductive}[\eta \mu ]\; \rbag
  \\& \hspace{1em}\;\ \!\centerdot\;    [[\; \texttt{ret}\; \cdotp \; (\texttt{lookup}\sharp[\; x\; ]\; \cdotp \; \rho \sharp)\; ]]
  \\& \hspace{1em}\;\ \!\centerdot\;    [[\; \texttt{pure}\; \cdotp \; \mathcal{A}\sharp[\; \texttt{Num}\; n\; ]\; \cdotp \; \rho \sharp\; ]]\;\;\square 
  \end{align*}
\vspace{-1em}\caption{Constructive GC calculations in Agda} \label{f:calc-agda} \vspace{-1em}\end{figure} 

\paragraph{The Precise Type System}

The AGT paper describes two designs for gradual type systems in increasing
complexity. We chose to mechanize a hybrid of the two which is simple, like the
first design, yet still exercises key challenges addressed by the second. We
also made slight modifications to the design at parts to make mechanization
easier, but without changing the nature of the system.

The precise type system we mechanized is the simply typed lambda calculus with
booleans, and top and bottom elements for a subtyping lattice, which we call
\texttt{any} and \texttt{none}:
\begin{align*}& \tau  \in  \texttt{type} \Coloneqq  \texttt{none} \mathrel{|} \mathbb{B} \mathrel{|} \tau  \rightarrow  \tau  \mathrel{|} \texttt{any} \end{align*}
The first design in the AGT paper does not involve subtyping, and their second
design incorporates record types with width and depth subtyping. By just
focusing on \texttt{none} and \texttt{any}, we exercise the subtyping machinery of their
approach without the blowup in complexity from formalizing record types.

The typing rules in AGT are written in strictly syntax-directed form, with
explicit use of subtyping in rule hypotheses. We show three precise typing
rules for if-statements, application and coercion in Figure~\ref{f:agt-precise}. The
subtyping lattice in the precise system is the ``safe for substitution'' lattice,
and well typed programs enjoy progress and preservation.

\paragraph{Gradual Types}

The essence of AGT is to design a gradual type system by \emph{abstract
interpretation} of the precise type system. To do this, a new top element is
added to the precise type system, although rather than representing the top of
the \emph{subtyping} lattice like \texttt{any}, it represents the top of the
\emph{precision} lattice, and is notated $\mathord{?}$:
\begin{align*}& \tau ^{\sharp } \in  \texttt{type}^{\sharp } \Coloneqq  \texttt{none} \mathrel{|} \mathbb{B} \mathrel{|} \tau ^{\sharp } \rightarrow  \tau ^{\sharp } \mathrel{|} \texttt{any} \mathrel{|} \mathord{?} \end{align*}
The partial ordering is reflexive and has $\mathord{?}$ at the top:
\begin{align*}& \tau ^{\sharp } \sqsubseteq  \tau ^{\sharp } \sqsubseteq  \mathord{?} \end{align*}
And arrow types are monotonic:
\begin{align*}& \tau _{11}^{\sharp } \sqsubseteq  \tau _{12}^{\sharp } \wedge  \tau _{21}^{\sharp } \sqsubseteq  \tau _{22}^{\sharp } \implies   \tau _{11}^{\sharp } \rightarrow  \tau _{21}^{\sharp } \sqsubseteq  \tau _{12}^{\sharp } \rightarrow  \tau _{22}^{\sharp } \end{align*}
Just as in our other designs by abstract interpretation, $\texttt{type}^{\sharp }$ is given
meaning by an interpretation function $\mu $, which is the constructive analog of
a classical concretization ($\gamma $) function:
\begin{align*}& \mu  : \texttt{type}^{\sharp } \shortnearrow  \wp (\texttt{type})
\\& \mu (\mathord{?}) \coloneqq  \{ \tau  \mathrel{|} \tau  \in  type\} 
\\& \mu (\tau _1 ^{\sharp } \rightarrow  \tau _2 ^{\sharp }) \coloneqq  \{ \tau _1  \rightarrow  \tau _2  \mathrel{|} \tau _1  \in  \mu (\tau _1 ^{\sharp }) \wedge  \tau _2  \in  \mu (\tau _2 ^{\sharp })\} 
\\& \mu (\tau ^{\sharp }) \coloneqq  \tau ^{\sharp } \quad \textit{w}\textit{h}\textit{e}\textit{n} \quad \tau ^{\sharp } \in  \{ \texttt{none},\mathbb{B},\texttt{any}\} 
\end{align*}
The extraction function $\eta $ is, remarkably, the identity function:
\begin{align*}& \eta  : \texttt{type}\! \rightarrow \! \texttt{type}^{\sharp } && \eta (\tau ) = \tau 
\end{align*}
and the constructive Galois correspondence holds:
\begin{align*}& \tau  \in  \mu (\tau ^{\sharp }) \iff   \eta (\tau ) \sqsubseteq  \tau ^{\sharp } \end{align*}

\paragraph{Gradual Operators}

Given the constructive Galois connection between gradual and precise types, we
synthesize specifications for abstract analogs of subtyping $<:$ and the
subtyping join operator $\overset{..}{\vee}$, and relate them to their abstractions $<:^{\sharp }$ and
$\overset{..}{\vee}^{\sharp }$:
\begin{align*}& \tau _1  \in  \mu (\tau ^{\sharp }_1 ) \wedge  \tau _2  \in  \mu (\tau ^{\sharp }_2 ) \wedge  \tau _1  <: \tau _2  \iff   \tau _1 ^{\sharp } <:^{\sharp } \tau ^{\sharp }_2 
\\& \langle \tau _1 ,\tau _2 \rangle  \in  \mu (\tau ^{\sharp }_1 ,\tau ^{\sharp }_2 ) \wedge  \tau ^{\sharp }_3  \sqsubseteq  \eta (\tau _1  \overset{..}{\vee} \tau _2 ) \iff   \tau ^{\sharp }_3  \sqsubseteq  \tau _1 ^{\sharp } \overset{..}{\vee}^{\sharp } \tau _2 ^{\sharp }
\end{align*}
Key properties of gradual subtyping and the gradual join operator is how they
operate over the unknown type $\mathord{?}$:
\begin{mathpar} \mathord{?} <:^{\sharp } \tau ^{\sharp }

 \tau ^{\sharp } <:^{\sharp } \mathord{?}

 \mathord{?} \overset{..}{\vee}^{\sharp } \tau ^{\sharp } = \tau ^{\sharp } \overset{..}{\vee}^{\sharp } \mathord{?} = \mathord{?}
\end{mathpar}

\begin{figure} 
  \begin{mathpar} \inferrule*[right=If]
     {\Gamma  \vdash  e_1  : \tau _1   \\ \tau _1  <: \mathbb{B}
  \\\\\Gamma  \vdash  e_2  : \tau _2   \\ \phantom{\tau _1  <: \mathbb{B}}
  \\\\\Gamma  \vdash  e_3  : \tau _3   \\ \phantom{\tau _1  <: \mathbb{B}} 
     }
     {\Gamma  \vdash  \texttt{if}\ e_1 \ \texttt{then}\ e_2 \ \texttt{else}\ e_3  : \tau _1  \overset{..}{\vee} \tau _2 }
  
 \inferrule*[right=App]
     {\Gamma  \vdash  e_1  : \tau _1   \\ \tau _1  <: \tau _{11} \rightarrow  \tau _{21}
  \\\\\Gamma  \vdash  e_2  : \tau _2   \\ \tau _2  <: \tau _{11} \phantom{\  \rightarrow  \tau _{21}}
     }
     {\Gamma  \vdash  e_1 (e_2 ) : \tau _{21}}
  
 \inferrule*[right=Coe]
     {\Gamma  \vdash  e : \tau _1  \\ \tau _1  <: \tau _2 }
     {\Gamma  \vdash  e ::  \tau _2  : \tau _2 }
  \end{mathpar}
\vspace{-1em}\caption{Case Study: precise type system} \label{f:agt-precise} \vspace{-1em}\end{figure} 

\paragraph{Gradual Metatheory}

Using AGT, the gradual type system is a syntactic analog to the precise one but
with gradual types and operators, which we show in Figure~\ref{f:agt-gradual}. Using
this system, and constructive Galois connections, we mechanize in Agda the key
AGT metatheory results from the paper: equivalence for fully-annotated terms
(FAT), embedding of dynamic language terms (EDL), and gradual guarantee (GG):
\begin{align*}& \vdash  e : \tau  \iff   \vdash ^{\sharp } e : \tau                                       \tag{FAT} 
\\& closed(un) \implies   \vdash ^{\sharp } \lceil un\rceil  : \mathord{?}                                    \tag{EDL}
\\& \vdash ^{\sharp } e^{\sharp }_1  : \tau ^{\sharp }_1 \ \wedge \ e^{\sharp }_1  \sqsubseteq  e^{\sharp }_2  \implies   \vdash ^{\sharp } e^{\sharp }_2  : \tau ^{\sharp }_2 \ \wedge \ \tau ^{\sharp }_1  \sqsubseteq  \tau ^{\sharp }_2  \tag{GG}
\end{align*}


\section{Constructive Galois Connection Metatheory} \label{s:metatheory} 

In this section we develop the full metatheory of constructive Galois
connection and prove precise claims about their relationship to classical
Galois connections. We develop the metatheory of constructive Galois
connections as an adjunction between posets with powerset-Kleisli adjoint
functors. This is in contrast to classical Galois connections which come from
an identical setup, but with the monotonic function space as adjoint functors,
as shown in Figure~\ref{f:table}.

\begin{figure} 
  \begin{center}
%
%
%
%
%

\def\firstcircle{(0,-.35) circle (1.2)}
\def\secondcircle{(0,0) circle (0.75)}
\def\thirdcircle{(2.5,0) circle (0.75)}
\def\fourthcircle{(5,0) circle (0.75)}

\colorlet{circle edge}{blue!50}
\colorlet{circle area}{blue!20}

\tikzset{filled/.style={fill=circle area, draw=circle edge, thick},
        outline/.style={draw=circle edge, thick}}


\begin{tikzpicture}[scale=1.04]
    \begin{scope}
      \clip \firstcircle;
    \end{scope}
    \draw[outline] \firstcircle  node {};
    \node[anchor=center] at (0,-1.1) {{\scriptsize Classical}};
    \begin{scope}
      \clip \secondcircle;
    \end{scope}
    \draw[outline] \secondcircle node {};
    \node[anchor=center] at (0,0) {{\scriptsize Computational}}; 
    \node at (0.60,0.20) {} edge[->,thick] ( 1.78,0.20);
    \node at ( 1.88,-0.2) {} edge[->,thick] (0.70,-0.2);
    \begin{scope}
      \clip \thirdcircle;
    \end{scope}
    \draw[outline] \thirdcircle node {};
    \node[anchor=center] at (2.5,0) {{\scriptsize Kleisli}};
    \node at (3.10,0.20) {} edge[->,thick,dotted] (4.28,0.20);
-   \node at (4.38,-0.2) {} edge[->,thick]        (3.20,-0.2);
    \begin{scope}
      \clip \fourthcircle;
    \end{scope}
    \draw[outline] \fourthcircle node {};
    \node[anchor=center] at (5,0) {{\scriptsize Constructive}};

\begin{customlegend}[
legend style={align=left,at={(5.7,-.9)},font=\scriptsize},
legend entries={ 
Set inclusion\hspace{2.3em},
Theorem of choice,
}
] 
    \addlegendimage{-stealth,black}
    \addlegendimage{-stealth,dotted}
\end{customlegend}
\end{tikzpicture}

  \end{center}
\vspace{-1em}\caption{Relationship between classical, Kleisli and constructive} \label{f:gcvenn} \vspace{-1em}\end{figure}
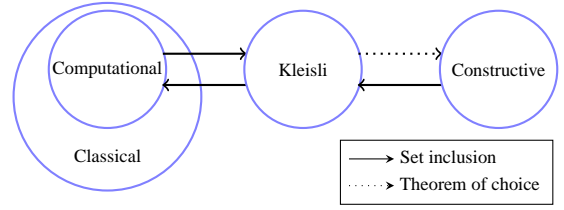

We connect constructive to classical Galois connections through an isomorphism
between a subset of classical to the entire space of constructive. To form this
isomorphism we introduce an intermediate structure, Kleisli Galois connections,
which we show are isomorphic to the classical subset, and isomorphic to
constructive ones using the constructive theorem of choice, as depicted in
Figure~\ref{f:gcvenn}.

\begin{figure} 
  \begin{mathpar} \inferrule*[right=G-If]
     {\Gamma ^{\sharp } \vdash ^{\sharp } e^{\sharp }_1  : \tau ^{\sharp }_1   \\ \tau ^{\sharp }_1  <:^{\sharp } \mathbb{B}
  \\\\\Gamma ^{\sharp } \vdash ^{\sharp } e^{\sharp }_2  : \tau ^{\sharp }_2   \\ \phantom{\tau ^{\sharp }_1  <:^{\sharp } \mathbb{B}}
  \\\\\Gamma ^{\sharp } \vdash ^{\sharp } e^{\sharp }_3  : \tau ^{\sharp }_3   \\ \phantom{\tau ^{\sharp }_1  <:^{\sharp } \mathbb{B}} 
     }
     {\Gamma ^{\sharp } \vdash ^{\sharp } \texttt{if}\ e_1 \ \texttt{then}\ e_2 \ \texttt{else}\ e_3  : \tau ^{\sharp }_1  \overset{..}{\vee}^{\sharp } \tau ^{\sharp }_2 }
  
 \inferrule*[right=G-App]
     {\Gamma ^{\sharp } \vdash ^{\sharp } e^{\sharp }_1  : \tau ^{\sharp }_1   \\ \tau ^{\sharp }_1  <:^{\sharp } \tau ^{\sharp }_{11} \rightarrow  \tau ^{\sharp }_{21}
  \\\\\Gamma ^{\sharp } \vdash ^{\sharp } e^{\sharp }_2  : \tau ^{\sharp }_2   \\ \tau ^{\sharp }_2  <:^{\sharp } \tau ^{\sharp }_{11} \phantom{\  \rightarrow  \tau ^{\sharp }_{21}}
     }
     {\Gamma ^{\sharp } \vdash  e^{\sharp }_1 (e^{\sharp }_2 ) : \tau ^{\sharp }_{21}}
  
 \inferrule*[right=G-Coe]
     {\Gamma ^{\sharp } \vdash ^{\sharp } e^{\sharp } : \tau ^{\sharp }_1  \\ \tau ^{\sharp }_1  <:^{\sharp } \tau ^{\sharp }_2 }
     {\Gamma ^{\sharp } \vdash ^{\sharp } e^{\sharp } ::  \tau ^{\sharp }_2  : \tau ^{\sharp }_2 }
  \end{mathpar}
\vspace{-1em}\caption{Case Study: gradual type system} \label{f:agt-gradual} \vspace{-1em}\end{figure} 

\paragraph{Classical Galois Connections}

We review classical Galois connections in Figure~\ref{f:table}. A Galois connection
between posets $A$ and $B$ contains two adjoint functors $\alpha $ and $\gamma $ which
share a correspondence. An equivalent formulation of the correspondence is two
unit equations called extensive and reductive. Abstract interpreters are sound
by over-approximating a specification induced by $\alpha $ and $\gamma $.

\paragraph{Powerset Monad}

See Sections~\ref{s:pos} and~\ref{s:spec-eff} for the downward-closure monotonicity
property, and monad definitions and notation for the monotonic powerset monad.
The monad operators obey standard monad laws. We introduce one new operator for
monadic function composition: $(g\circledast f)(x) \coloneqq  g^*(f(x))$.

\paragraph{Kleisli Galois Connections}

We summarize Kleisli Galois connections in Figure~\ref{f:table}. Kleisli Galois
connections are analogous to classical ones, but with monadic analogs to $\alpha $
and $\gamma $, and monadic identity and composition operators $ret$ and $\circledast $ in place
of the function space identity and composition operators $id$ and $\circ $. 

\paragraph{Kleisli to Classical and Back}

All Kleisli Galois connections $\langle \kappa \alpha ,\kappa \gamma \rangle $ between $A$ and $B$ can be lifted to
recover a classical Galois connection $\langle \alpha ,\gamma \rangle $ between $\wp (A)$ and $\wp (B)$ through
a monadic lifting operator on Kleisli Galois connections $\langle \kappa \alpha ,\kappa \gamma \rangle ^*$:
\begin{align*}& \langle \alpha ,\gamma \rangle  = \langle \kappa \alpha ,\kappa \gamma \rangle ^* = \langle \kappa \alpha ^*,\kappa \gamma ^*\rangle  \end{align*}
This lifting is \emph{sound}, meaning Kleisli soundness and optimality
results can be translated to classical ones.
\begin{theorem}[KGC-Sound]\trueaf
  For any Kleisli relationship of soundness between $f$ and $f^{\sharp }$, that is $\kappa \alpha  \circledast 
  f \circledast  \kappa \gamma  \sqsubseteq  f^{\sharp }$, its lifting to classical is also sound, that is $\alpha  \circ  f^* \circ  \gamma 
  \sqsubseteq  f^{\sharp *}$ where $\langle \alpha ,\gamma \rangle  = \langle \kappa \alpha ,\kappa \gamma \rangle ^*$, and likewise for optimality relationships.
\end{theorem}

This lifting is also \emph{complete}, meaning classical Galois connection
soundness and optimality results can always be translated to Kleisli ones, when
$\alpha $ and $\gamma $ are of lifted form.
\begin{theorem}[KGC-Complete]\trueaf
  For any classical relationship of soundness between $f^*$ and $f^{\sharp *}$, that is $\alpha 
  \circ  f^* \circ  \gamma  \sqsubseteq  f^{\sharp *}$, its lowering to Kleisli is also sound when $\langle \alpha ,\gamma \rangle  =
  \langle \kappa \alpha ,\kappa \gamma \rangle ^*$, that is $\kappa \alpha  \circledast  f \circledast  \kappa \gamma  \sqsubseteq  f^{\sharp }$, and likewise for optimality
  relationships.
\end{theorem}

Due to soundness and completeness, one can work with the simpler setup of
Kleisli Galois connections without any loss of generality. The setup is simpler
because Kleisli Galois connection theorems only quantify over individual
elements rather than elements of powersets. For example, the soundness criteria
$\kappa \alpha  \circledast  f \circledast  \kappa \gamma  \sqsubseteq  f^{\sharp }$ is proved by showing $\kappa \alpha ^*(f^*(\kappa \gamma (x))) \subseteq  f^{\sharp }(x)$ for an
arbitrary element $x : A$, whereas in the classical proof one must show
$\kappa \alpha ^*(f^*(\kappa \gamma ^*(X))) \subseteq  f^{\sharp *}(X)$ for arbitrary sets $X : \wp (A)$.

\begin{figure} 
  \begin{center}
  \begin{tabular}{ | r | l | l | }
    \hline \textit{Adjunction}   & Classical GCs      & Kleisli/constructive GCs 
  \\\hline \textit{Category}     & Posets             & Posets
  \\\hline \textit{Adjoints}     & Mono. Functions    & $\wp $-Monadic Functions
  \\\hline \textit{LAdjoint}     & $\alpha  : A \shortnearrow  B$        & $\kappa \alpha  : A \shortnearrow  \wp (B)$
  \\\hline \textit{RAdjoint}     & $\gamma  : B \shortnearrow  A$        & $\kappa \gamma  : B \shortnearrow  \wp (A)$
  \\\hline \textit{Corr}         & $id(x) \sqsubseteq  \gamma (y)$     & $ret(x) \subseteq  \kappa \gamma (y) $
  \\                             & $\Leftrightarrow  \alpha (x) \sqsubseteq  id(y)$   & $\Leftrightarrow  \kappa \alpha (x) \subseteq  ret(y)$
  \\\hline \textit{Extensive}    & $id \sqsubseteq  \gamma  \circ  \alpha $       & $ret \sqsubseteq  \kappa \gamma  \circledast  \kappa \alpha $
  \\\hline \textit{Reductive}    & $\alpha  \circ  \gamma  \sqsubseteq  id$       & $\kappa \alpha  \circledast  \kappa \gamma  \sqsubseteq  ret$
  \\\hline \textit{Soundness}    & $\alpha  \circ  f \circ  \gamma  \sqsubseteq  f^{\sharp }$   & $\kappa \alpha  \circledast  f \circledast  \kappa \gamma  \sqsubseteq  f^{\sharp }$
  \\\hline \textit{Optimality}   & $\alpha  \circ  f \circ  \gamma  = f^{\sharp }$   & $\kappa \alpha  \circledast  f \circledast  \kappa \gamma  = f^{\sharp }$
  \\\hline \end{tabular}
  \end{center}
\vspace{-1em}\caption{Comparison of constructive v classical adjunctions} \label{f:table} \vspace{-1em}\end{figure} 

\paragraph{Constructive Galois Connections}

Constructive Galois connections are a restriction of Kleisli Galois connections
where the abstraction mapping is a pure rather than monadic function. We call
the left adjoint \emph{extraction}, notated $\eta $, and the right adjoint
\emph{interpretation}, notated $\mu $. The constructive Galois connection
correspondence, alternative expansive and reductive formulation of the
correspondence, and soundness and optimality criteria are identical to Kleisli
Galois connections where $\langle \kappa \alpha ,\kappa \gamma \rangle  = \langle pure(\eta ),\mu \rangle $.

\paragraph{Constructive to Kleisli and Back}

Our main theorem which justifies the soundness and completeness of constructive
Galois connections is an isomorphism between constructive and Kleisli Galois
connections. The easy direction is soundness, where a Kleisli Galois connection
is formed by defining $\langle \kappa \alpha ,\kappa \gamma \rangle  = \langle pure(\eta ),\mu \rangle $. Soundness and optimality
theorems are then lifted from constructive to Kleisli without modification.
\begin{theorem}[CGC-Sound]\trueaf
  For any constructive relationship of soundness between $f$ and $f^{\sharp }$, that is
  $pure(\eta ) \circledast  f \circledast  \mu  \sqsubseteq  f^{\sharp }$, its lifting to classical is sound, that is $\kappa \alpha  \circledast  f \circledast 
  \kappa \gamma  \sqsubseteq  f^{\sharp }$ where $\langle \kappa \alpha ,\kappa \gamma \rangle \! =\! \langle pure(\eta ),\mu \rangle $, and likewise for optimality relationships.
\end{theorem}
The other direction, completeness, is much more surprising. First we establish
a lowering for Kleisli Galois connections.
\begin{lemma}[CGC-Induce]\trueaf
  For every Kleisli Galois connection $\langle \kappa \alpha ,\kappa \gamma \rangle $, there exists a constructive
  Galois connection $\langle \eta ,\mu \rangle $ where $\langle pure(\eta ),\mu \rangle  = \langle \kappa \alpha ,\kappa \gamma \rangle $.
\end{lemma}
Because the mapping from Kleisli to constructive is interesting we provide a
proof, which to our knowledge is novel. The proof builds a constructive Galois
connection $\langle \eta ,\mu \rangle $ from a Kleisli $\langle \kappa \alpha ,\kappa \gamma \rangle $ by exploiting the Kleisli
correspondence and making use of the constructive theorem of choice.

\begin{proof}

To turn an arbitrary Kleisli Galois connection into a constructive one, we show
that the effect on $\kappa \alpha  : A \shortnearrow  \wp (B)$ is benign, or in other words, that there
exists some $\eta $ such that $\kappa \alpha =pure(\eta )$. We prove this using two ingredients: a
constructive interpretation of the Kleisli extensive law, and the constructive
\emph{theorem} of choice.

We first expand the Kleisli expansive property, unfolding definitions of $\circledast $
and $ret$, to get an equivalent logical statement:
\begin{align*}& \forall x. \exists y. y \in  \kappa \alpha (x) \wedge  x \in  \kappa \gamma (y) \tag{KGC-Exp} \label{e:KGC-Exp} \end{align*}
Statements of this form can be used in conjunction with an axiom of choice in
classical mathematics, which is:
\begin{align*}& (\forall x.\exists y.R(x,y)) \implies   (\exists f.\forall x.R(x,f(x))) \tag{AxChoice} \label{e:Choice} \end{align*}
This theorem is admitted as an \emph{axiom} in classical mathematics, but in
constructive logic---the setting used for extracting verified
algorithms--\eqref{e:Choice}is definable as a \emph{theorem}, due to the computational
interpretation of logical connectives $\forall $ and $\exists $. We define \eqref{e:Choice} as a
theorem in Agda without trouble:
\begin{align*}& \texttt{choice} : \forall  \{ A\ B\}  \{ R : A \rightarrow  B \rightarrow  \texttt{Set}\}  
\\& \rightarrow  (\forall \ x \rightarrow  \exists \ y\ \textit{s}\textit{t}\ R\ x\ y) 
\\& \rightarrow  (\exists \ f\ \textit{s}\textit{t}\ \forall \ x \rightarrow  R\ x\ (f\ x))
\\& \texttt{choice}\ P = \exists \ (\lambda \ x \rightarrow  \pi _1 \ (P\ x))\ ,\ (\lambda \ x \rightarrow  \pi _2 \ (P\ x)) 
\end{align*}
Applying \eqref{e:Choice} to \eqref{e:KGC-Exp} then gives:
\begin{align*}& \exists \eta . \forall x. \eta (x) \in  \kappa \alpha (x) \wedge  x \in  \kappa \gamma (\eta (x)) \tag{ExpChioce} \label{e:ExpChoice} \end{align*}
which proves the existence of a pure function $\eta  : A \shortnearrow  B$.

In order to form a constructive Galois connection $\eta $ and $\mu $ must satisfy the
correspondence, which we prove in split form:
\begin{align*}& x \in  \mu (\eta (x))                 \tag{CGC-Exp}
\\& x \in  \mu (y) \implies   \eta (x) \sqsubseteq  y        \tag{CGC-Red}
\end{align*}
The expansive property is immediate from the second conjunct in \eqref{e:ExpChoice}.
The reductive property follows from the Kleisli reductive property:
\begin{align*}& x \in  \kappa \gamma (y) \wedge  y' \in  \kappa \alpha (x) \implies   y' \sqsubseteq  y \tag{KGC-Red} \label{e:KGC-Red} \end{align*}
The constructive variant of reductive is proved by satisfying the
first two premises of \eqref{e:KGC-Red}, where $x \in  \kappa \gamma (y)$ is by assumption and $y' \in 
\kappa \alpha (x)$ is by the first conjunct in \eqref{e:ExpChoice}.

So far we have shown that for a Kleisli Galois connection $\langle \kappa \alpha ,\kappa \gamma \rangle $, there
exists a constructive Galois connection $\langle \eta ,\mu \rangle $ where $\mu =\kappa \gamma $. However, we have
yet to show $pure(\eta )=\kappa \alpha $. To show this, we prove an analog of a standard result
for classical Galois connections: that $\alpha $ and $\gamma $ uniquely determine each
other.
\begin{lemma}[Unique Abstraction]\trueaf
  For any two Kleisli Galois connections $\langle \kappa \alpha _1 ,\kappa \gamma _1 \rangle $ and $\langle \kappa \alpha _2 ,\kappa \gamma _2 \rangle $, $\kappa \alpha _1 \!=\!\kappa \alpha _2 $
  \textit{i}\textit{f}\textit{f} $\kappa \gamma _1 =\kappa \gamma _2 $
\end{lemma}
We then conclude $pure(\eta )=\kappa \alpha $ as a consequence of the above lemma and the fact
that $\mu =\kappa \gamma $.

\end{proof}

Given the above mapping from Kleisli Galois connections to constructive ones,
we prove the completeness of this mapping.
\begin{theorem}[CGC-Complete]\trueaf
  For any Kleisli relationship of soundness between $f$ and $f^{\sharp }$, that is $\kappa \alpha  \circledast 
  f \circledast  \kappa \gamma  \sqsubseteq  f^{\sharp }$, its lowering to constructive is also sound, that is $pure(\eta ) \circledast 
  f \circledast  \mu  \sqsubseteq  f^{\sharp }$ where $\langle \eta ,\mu \rangle $ is induced, and likewise for optimality
  relationships.
\end{theorem}

\paragraph{Mechanization}

We mechanize the metatheory for constructive Galois connections and both case
studies from Sections~\ref{s:case-cdgai} and~\ref{s:case-agt} in Agda, as well as a
general purpose proof library for posets and calculational reasoning with the
monotonic powerset monad. The development is available at:
\texttt{github.com/plum-umd/cgc}.

\paragraph{Wrapping Up}

In this section we showed that constructive Galois connections are sound w.r.t.
classical Galois connections, and complete w.r.t. the subset of classical
Galois connections recovered by lifting constructive ones. We showed this
by introducing an intermediate space of Galois connections, Kleisli Galois
connections, and by establishing two sets of isomorphisms between a subset of
classical and Kleisli, and between Kleisli and constructive. The proof of
isomorphism between constructive and Kleisli yielded an interesting proof which
applies the constructive \emph{theorem} of choice to one of the Kleisli Galois
connection correspondence laws.


\section{Related Work} \label{s:related} 

This work connects two long strands of research: abstract interpretation via
Galois connections and mechanized verification via dependently typed functional
programming.  The former is founded on the pioneering work of Cousot and
Cousot~\cite{dvanhorn:Cousot:1977:AI,dvanhorn:Cousot1979Systematic}; the latter
on that of Martin-L\"of~\cite{local:lof}, embodied in Norell's
Agda~\cite{local:norell:thesis}.  Our key technical insight is to use a monadic
structure for Galois connections, following the example of
Moggi~\cite{davdar:Moggi:1989:Monads} for the $\lambda$-calculus.

\paragraph{Calculational Abstract Interpretation}

Cousot describes calculational abstract interpretation
by example in his lecture notes~\cite{local:cousot-mit} and
monograph~\cite{davdar:cousot:1999:calculational}, and recently
introduced a unifying calculus for Galois
connections~\cite{dvanhorn:Cousot2014Galois}.
Our work mechanizes Cousot's calculations and provides a
foundation for mechanizing other instances of calculational abstract
interpretation (e.g.~\cite{davdar:midtgaard:2008:calculational-cfa,
  dvanhorn:Sergey2012Calculating}).
We expect our work to have applications to the mechanization of
calculational program
design~\cite{local:algebra-of-programming,local:Bird90:Calculus} by
employing only Galois \emph{retractions}, i.e.~$\alpha \circ \gamma$
is an identity~\cite{dvanhorn:Cousot2014Galois}.  There is prior work
on mechanized program calculation~\cite{dvanhorn:Tesson2011Program},
but it is not based on abstract interpretation.

\paragraph{Verified Static Analyzers}

Verified abstract interpretation has many promising
results~\cite{local:pichardie:2005:phd-thesis,dvanhorn:Cachera2010Certified,dvanhorn:Blazy2013Formal,dvanhorn:Barthe2007Certified},
scaling up to large-scale real-world static
analyzers~\cite{dvanhorn:Jourdan2015FormallyVerified}.
However, mechanized abstract interpretation has yet to benefit from the Galois
connection framework. Until now, approaches use classical axioms or
``$\gamma $-only'' encodings of soundness and (sometimes) completeness. Our
techniques for mechanizing Galois connections should complement these
approaches.

\paragraph{Galculator}

The \emph{Galculator}~\cite{dvanhorn:Silva2008Galculator} is a proof assistant
founded on an algebra of Galois connections.  This tool is similar to ours in
that it mechanically verifies Galois connection calculations. Our approach is
more general, supporting arbitrary set-theoretic reasoning and embedded within
a general purpose proof assistant, however their approach is fully automated
for the small set of derivations which reside within their supported theory. 

\paragraph{Deductive Synthesis}
Fiat~\cite{local:fiat} is a library for the Coq proof assistant which supports
semi-automated synthesis of programs as refinements of their specifications.
Fiat uses the same powerset type and monad as we do, and their ``deductive
synthesis'' process similarly derives correct-by-construction programs by
calculus. Fiat derivations start with a user-defined specification and
calculate towards an \emph{under}-approximation ($\sqsupseteq $), whereas calculational
abstract interpretation starts with an optimal specification and calculates
towards an \emph{over}-approximation ($\sqsubseteq $). It should be possible to generalize
their framework to use partial orders to recover aspects of our work, or to
invert the lattice used in our abstract interpretation framework to recover
aspects of theirs. A notable difference in approach is that Fiat makes heavy
use of Coq's tactic programming language to automate rewrites inside respectful
contexts, whereas our system provides no interactive proof automation and each
calculational step must be justified explicitly.

\paragraph{Monadic Abstract Interpretation}

Monads in abstract interpretation have recently been applied to good effect for
modularity~\cite{dvanhorn:Sergey2013Monadic, Darais:2015:GTM:2814270.2814308}.
However, that work uses monads to structure the semantics, not the Galois
connections and proofs.

\paragraph{Future Directions}

Now that we have established a foundation for constructive Galois
connection calculation, we see value in verifying larger derivations
(e.g.~\cite{dvanhorn:midtgaard-jensen-sas-08,
  dvanhorn:Sergey2012Calculating}).
Furthermore we would like to explore whether or not our techniques
have any benefit in the space of general-purpose program calculations
\emph{\`a la} Bird.

Currently our framework requires the user to justify every detail of the
program calculation, including monotonicity proofs and proof scoping for
rewrites inside monotonic contexts. We imagine much of this can be automated,
requiring the user to only provide the interesting parts of the proof, \`a la
Fiat\cite{local:fiat}. Our experience has been that even Coq's tactic system
slows down considerably when automating all of these details, and we foresee
using proof by reflection in either Coq (\emph{e.g.} Rtac~\cite{local:rtac}) or
Agda to automate these proofs in a way that maintains proof-checker
performance.

There have been recent developments on compositional abstract
interpretation frameworks~\cite{Darais:2015:GTM:2814270.2814308} where
abstract interpreters and their proofs of soundness are systematically
derived side-by-side. That framework relies on correctness properties
transported by \emph{Galois transformers}, which we posit would
benefit from mechanization since they hold both computational and
specification content.


\section{Conclusions} \label{s:conclusions} 

This paper realizes the vision of mechanized and constructive Galois
connections foreshadowed by
Cousot~\cite[p.~85]{davdar:cousot:1999:calculational}, giving the
first mechanically verified proof by calculational abstract
interpretation; once for his generic static analyzer and once for
the semantics of gradual typing.  Our proofs by calculus closely
follow the originals.  The primary discrepancy is the use of
monads to isolate \emph{specification effects}.  By
maintaining this discipline, we are able to verify calculations by
Galois connections \emph{and} extract computational content from pure
results.  The resulting artifacts are
correct-by-verified-construction, thereby avoiding known
bugs in the original.\footnote{\scriptsize\href{http://www.di.ens.fr/~cousot/aisoftware/Marktoberdorf98/Bug_History}{\tt di.ens.fr/\textasciitilde cousot/aisoftware/Marktoberdorf98/Bug\_History}}

\acks{We thank Ron Garcia and \'Eric Tanter for discussions of their work.
\'Eric also helped with our French translation.  We thank the Colony Club in
D.C.~and the Board \& Brew in College Park for providing fruitful environments
in which to work. We thank the anonymous reviewers of ICFP 2016 for their
helpful feedback. This material is partially based on research sponsored by
DARPA under agreement number AFRL FA8750-15-2-0104.
}


\balance
\bibliographystyle{abbrvnat}
\bibliography{davdar,dvanhorn,local}

%

\end{document}